\input harvmac
\input epsf.tex
\def\N{{\cal N}}

\def\nc{N_c}
\def\nf{N_f}
\def\ncd{\widetilde{N_c}}

\def\yt{\tilde y}
\def\xt{\tilde x}
\def\nt{\tilde n}
\def\at{\tilde a}
\noblackbox
%
% dual defs
%

\lref\GW{D.~J.~Gross and F.~Wilczek,
``Asymptotically Free Gauge Theories. 2,''
Phys.\ Rev.\ D {\bf 9}, 980 (1974).
%%CITATION = PHRVA,D9,980;%%
}
%\PoppitzVH
\lref\PoppitzVH{
E.~Poppitz, Y.~Shadmi and S.~P.~Trivedi,
``Duality and Exact Results in Product Group Theories,''
Nucl.\ Phys.\ B {\bf 480}, 125 (1996)
[arXiv:hep-th/9605113].
%%CITATION = HEP-TH 9605113;%%
}

%\MachacekTZ
\lref\twoloop{
M.~E.~Machacek and M.~T.~Vaughn,
``Two Loop Renormalization Group Equations In A General Quantum Field Theory.
1. Wave Function Renormalization,''
Nucl.\ Phys.\ B {\bf 222}, 83 (1983).
%%CITATION = NUPHA,B222,83;%%
}

\lref\BZ{T.~Banks and A.~Zaks,
``On The Phase Structure Of Vector - Like Gauge Theories With Massless
Fermions,''
Nucl.\ Phys.\ B {\bf 196}, 189 (1982).
%%CITATION = NUPHA,B196,189;%%
}

\lref\Cardy{
J.~L.~Cardy,
``Is There A C Theorem In Four-Dimensions?,''
Phys.\ Lett.\ B {\bf 215}, 749 (1988).
%%CITATION = PHLTA,B215,749;%%
}

\lref\Zam{A.~B.~Zamolodchikov,
``'Irreversibility' Of The Flux Of The Renormalization Group In A 2-D
Field Theory,''
JETP Lett.\  {\bf 43}, 730 (1986)
[Pisma Zh.\ Eksp.\ Teor.\ Fiz.\  {\bf 43}, 565 (1986)].
%%CITATION = JTPLA,43,730;%%
}

\lref\NSVZ{V.~A.~Novikov, M.~A.~Shifman, A.~I.~Vainshtein and
V.~I.~Zakharov,
``Exact Gell-Mann-Low Function Of Supersymmetric Yang-Mills Theories
From
Instanton Calculus,''
Nucl.\ Phys.\ B {\bf 229}, 381 (1983).
%%CITATION = NUPHA,B229,381;%%
}

\lref\DFHO{
D.~Z.~Freedman and H.~Osborn,
``Constructing a c-function for SUSY gauge theories,''
Phys.\ Lett.\ B {\bf 432}, 353 (1998)
[arXiv:hep-th/9804101].
%%CITATION = HEP-TH 9804101;%%
}

\lref\osborn{
H.~Osborn,
``Derivation Of A Four-Dimensional C Theorem,''
Phys.\ Lett.\ B {\bf 222}, 97 (1989);
%%CITATION = PHLTA,B222,97;%%
H.~Osborn,
``Weyl Consistency Conditions And A Local Renormalization Group Equation For
General Renormalizable Field Theories,''
Nucl.\ Phys.\ B {\bf 363}, 486 (1991);
%%CITATION = NUPHA,B363,486;%%}
I.~Jack and H.~Osborn,
``Analogs For The C Theorem For Four-Dimensional Renormalizable Field
Theories,''
Nucl.\ Phys.\ B {\bf 343}, 647 (1990).
%%CITATION = NUPHA,B343,647;%%
}
%\KutasovXU
\lref\DKASlm{
D.~Kutasov and A.~Schwimmer,
``Lagrange Multipliers and Couplings in Supersymmetric Field Theory,''
arXiv:hep-th/0409029.
%%CITATION = HEP-TH 0409029;%%
}

\lref\NSd{N.~Seiberg,
``Electric - magnetic duality in supersymmetric nonAbelian
gauge theories,''Nucl.\ Phys.\ B {\bf 435}, 129
(1995)[arXiv:hep-th/9411149].
%%CITATION = HEP-TH 9411149;%%}
}

%\IntriligatorAU
\lref\ISrev{
K.~A.~Intriligator and N.~Seiberg,
``Lectures on supersymmetric gauge theories and electric-magnetic  duality,''
Nucl.\ Phys.\ Proc.\ Suppl.\  {\bf 45BC}, 1 (1996)
[arXiv:hep-th/9509066].
%%CITATION = HEP-TH 9509066;%%
}

\lref\KINS{K.~A.~Intriligator and N.~Seiberg,
``Phases of N=1 supersymmetric gauge theories in four-dimensions,''
Nucl.\ Phys.\ B {\bf 431}, 551 (1994)
[arXiv:hep-th/9408155].
%%CITATION = HEP-TH 9408155;%%
}

\lref\DKi{
D.~Kutasov,
``A Comment on duality in N=1 supersymmetric nonAbelian gauge
theories,''
Phys.\ Lett.\ B {\bf 351}, 230 (1995)
[arXiv:hep-th/9503086].
%%CITATION = HEP-TH 9503086;%%
}

\lref\DFHO{
D.~Z.~Freedman and H.~Osborn,
``Constructing a c-function for SUSY gauge theories,''
Phys.\ Lett.\ B {\bf 432}, 353 (1998)
[arXiv:hep-th/9804101].
%%CITATION = HEP-TH 9804101;%%
}

%\KutasovUX
\lref\DKlm{
D.~Kutasov,
 ``New results on the 'a-theorem' in four dimensional supersymmetric field
theory,''
arXiv:hep-th/0312098.
%%CITATION = HEP-TH 0312098;%%
}

%\IntriligatorMI
\lref\IWii{
K.~Intriligator and B.~Wecht,
``RG fixed points and flows in SQCD with adjoints,''
Nucl.\ Phys.\ B {\bf 677}, 223 (2004)
[arXiv:hep-th/0309201].
%%CITATION = HEP-TH 0309201;%%
}

\lref\DKAS{
D.~Kutasov and A.~Schwimmer,
``On duality in supersymmetric Yang-Mills theory,''
Phys.\ Lett.\ B {\bf 354}, 315 (1995)
[arXiv:hep-th/9505004].
%%CITATION = HEP-TH 9505004;%%
}

\lref\DKNSAS{D.~Kutasov, A.~Schwimmer and N.~Seiberg,
``Chiral Rings, Singularity Theory and Electric-Magnetic Duality,''
Nucl.\ Phys.\ B {\bf 459}, 455 (1996)
[arXiv:hep-th/9510222].
%%CITATION = HEP-TH 9510222;%%
}

%\IntriligatorAX
\lref\ILS{
K.~A.~Intriligator, R.~G.~Leigh and M.~J.~Strassler,
 ``New examples of duality in chiral and nonchiral supersymmetric gauge
theories,''
Nucl.\ Phys.\ B {\bf 456}, 567 (1995)
[arXiv:hep-th/9506148].
%%CITATION = HEP-TH 9506148;%%
}

%\IntriligatorFF
\lref\kispso{
K.~A.~Intriligator,
``New RG fixed points and duality in supersymmetric SP(N(c)) and SO(N(c)) gauge
theories,''
Nucl.\ Phys.\ B {\bf 448}, 187 (1995)
[arXiv:hep-th/9505051].
%%CITATION = HEP-TH 9505051;%%
}

\lref\Luty{
M.~A.~Luty, M.~Schmaltz and J.~Terning,
``A Sequence of Duals for Sp(2N) Supersymmetric Gauge Theories with Adjoint
Matter,''
Phys.\ Rev.\ D {\bf 54}, 7815 (1996)
[arXiv:hep-th/9603034].
%%CITATION = HEP-TH 9603034;%%
}
%\BerkoozKM
\lref\Berkooz{
M.~Berkooz,
``The Dual of supersymmetric SU(2k) with an antisymmetric tensor and composite
dualities,''
Nucl.\ Phys.\ B {\bf 452}, 513 (1995)
[arXiv:hep-th/9505067].
%%CITATION = HEP-TH 9505067;%%
}

%\BrodieVV
\lref\BCI{
J.~H.~Brodie, P.~L.~Cho and K.~A.~Intriligator,
``Misleading anomaly matchings?,''
Phys.\ Lett.\ B {\bf 429}, 319 (1998)
[arXiv:hep-th/9802092].
%%CITATION = HEP-TH 9802092;%%
}

%\IntriligatorRX
\lref\ISS{
K.~A.~Intriligator, N.~Seiberg and S.~H.~Shenker,
``Proposal for a simple model of dynamical SUSY breaking,''
Phys.\ Lett.\ B {\bf 342}, 152 (1995)
[arXiv:hep-ph/9410203].
%%CITATION = HEP-PH 9410203;%%
}

%\OsbornAZ
\lref\OsbornAZ{
H.~Osborn and G.~M.~Shore,
``Correlation functions of the energy momentum tensor on spaces of  constant
curvature,''
Nucl.\ Phys.\ B {\bf 571}, 287 (2000)
[arXiv:hep-th/9909043].
%%CITATION = HEP-TH 9909043;%%
}

%\LeighEP
\lref\rlmsspso{
R.~G.~Leigh and M.~J.~Strassler,
``Exactly marginal operators and duality in four-dimensional N=1 supersymmetric
gauge theory,''
Nucl.\ Phys.\ B {\bf 447}, 95 (1995)
[arXiv:hep-th/9503121].
%%CITATION = HEP-TH 9503121;%%
}

\def\kdual{\refs{\DKi, \DKAS, \DKNSAS}}

\

%  draw box of size #1pt and line thickness #2pt
\def\drawbox#1#2{\hrule height#2pt
             \hbox{\vrule width#2pt height#1pt \kern#1pt \vrule
width#2pt}
                   \hrule height#2pt}
% Young tableaux

\def\Fund#1#2{\vcenter{\vbox{\drawbox{#1}{#2}}}}
\def\Asym#1#2{\vcenter{\vbox{\drawbox{#1}{#2}
                   \kern-#2pt       % line up boxes
                   \drawbox{#1}{#2}}}}

\def\fund{\Fund{6.5}{0.4}}
\def\asym{\Asym{6.5}{0.4}}
\def\sym{\Fund{6.5}{0.4} \kern-.5pt \Fund{6.5}{0.4}}

\Title{\vbox{\baselineskip12pt\hbox{hep-th/0502049}
\hbox{UCSD-PTH-04-10}
\hbox{MIT-CTP-3596} }}
{\vbox{\centerline{$\N =1$ RG Flows, Product Groups, and a-Maximization}}}
\centerline{ Edwin Barnes$^1$, Ken Intriligator$^1$,
Brian Wecht$^2$, and Jason Wright$^1$}
\bigskip
\centerline{$^1$Department of Physics} \centerline{University of
California, San Diego} \centerline{La Jolla, CA 92093-0354, USA}
\bigskip
\centerline{$^2$Center for Theoretical Physics} 
\centerline{Massachusetts Institute of Technology}
\centerline{Cambridge, MA 02139}

\bigskip
\noindent
We explore new IR phenomena and dualities, arising for product groups, in the context of 
$\N =1$ supersymmetric gauge theories.  The RG running of the multiple couplings can radically
affect each other.  For example, an otherwise IR interacting coupling can be driven to be
instead IR free by an arbitrarily small, but non-zero, initial value of another coupling.  Or an
otherwise IR free coupling can be driven to be instead IR interacting by  an arbitrarily small  non-zero initial value of another coupling. We explore these and other phenomena in $\N =1$ examples, where exact results can be obtained using a-maximization. 
We also explore the various possible dual gauge theories, 
e.g. by dualizing one gauge group with the other treated as a weakly gauged flavor symmetry, along with previously proposed duals for
the theories deformed by $A_k$-type Landau-Ginzburg superpotentials.  We note that this latter duality, and all similar duality examples, always have non-empty superconformal windows within
which both the electric and dual $A_k$ superpotentials are relevant.

%\draftmode
\Date{February  2005}
\newsec{Introduction}

Asymptotically free gauge theories have various possible IR phases, one being the
the ``non-Abelian Coulomb phase," which is an interacting conformal field theory RG fixed
point, where  all beta functions vanish.  A classic example is $\N =1$ $SU(N_c)$ SQCD with
$N_f$ massless flavors, which flows to a SCFT in the IR for $N_f$ within the Seiberg 
superconformal window \NSd\ ${3\over 2}N_c<N_f<3N_c$.  For $N_f\leq {3\over 2}N_c$, the 
theory is instead in a free magnetic $SU(N_c-N_f)$ phase in the IR.  
(See e.g. \ISrev\ for a review and references.)  Our prejudice is that the interacting
SCFT phase is rather generic for asymptotically free SUSY gauge theories with enough massless matter to avoid dynamical superpotentials, i.e. with massless matter representation
$R$ such that $T(G)<T(R)<3T(G)$, with $T(R)$ the quadratic Casimir of $R$ and $T(G)$ that of the adjoint.   The theory at the origin is then either a non-trivial 
free field solution of 't Hooft anomaly matching (as in the free magnetic phase) or an interacting SCFT.  Unfortunately, unless one has a conjectured dual description\foot{Even with a non-trivial
free field solution of 't Hooft anomaly matching, e.g. as in the example of \ISS, there's the possibility that the matching is a fluke, and that the theory actually flows to an interacting SCFT after all, as was argued to be the case for another example in \BCI.}, there is no simple test for directly determining if the IR phase is a SCFT or (fully or partially) IR free magnetic.

There is an essentially endless landscape of possible RG fixed point SCFTs to explore, coming {}from various gauge groups, including product groups, and matter representations.
 Here we'll consider examples with product gauge groups, e.g. the theory
\eqn\sunsun{\matrix{&&\hbox{gauge group:} &SU(N_c)\times SU(N_c')&\cr
&\hbox{matter:} &X \oplus \widetilde X&\bf{(\fund , \overline{\fund} )}\oplus \bf{(\overline{\fund},
\fund )},&\cr
&&Q_f \oplus \widetilde Q_{\tilde f}&\bf{(\fund , 1)} \oplus \bf{(\overline{\fund}, 1)}&(f, \widetilde f=1\dots N_f),\cr
&&Q'_{f'}\oplus \widetilde Q'_{\tilde f'}& \bf{(1, \fund )}\oplus \bf{(1, \overline{\fund})} &(f',\widetilde f'=1\dots N_f'). \cr}}
We'll be interested in when this theory flows to an (either fully or partially) interacting SCFT and
when various dualities apply, e.g. dualizing one gauge group with the other treated as a spectator.
We'll also be interested in the superconformal window for a duality proposed in \ILS, for the above  theory deformed by superpotential $W_{A_{2k+1}}=\Tr (X\widetilde X)^{k+1}$.

With multiple couplings, e.g. the two gauge couplings of \sunsun, the RG running
of one coupling can radically affect that of the other, possibly driving it into another basin of 
attraction.  For example, as depicted in fig. 1, there can be saddle point IR fixed points (A) and (B) when one or the other coupling is tuned to precisely zero, but which are unstable to any perturbation in the other coupling: the generic RG flows then end up at point (C) in the IR, with both $g_*$ and $g'_*$ nonzero.   Another possibility, shown in fig. 2, is that $g'$ is interacting in the IR only if $g=0$, but
that any arbitrarily small, nonzero,  $g$ would 
eventually overwhelm $g'$, and drive $g'$ to be IR irrelevant, $g'\rightarrow 0$ in the IR;
generic RG flows then end up at point (A), with $g'_*=0$.
 Fig. 3 depicts
an opposite situation, where an otherwise IR free coupling $g'$ is driven to be interacting
in the IR by the coupling $g$.  Fig. 4 depicts two separately IR free couplings, which can cure each
other and lead to an interacting RG fixed point (this happens for e.g. the gauge and Yukawa couplings
of the $\N=4$ theory, when we break to $\N =1$ by taking them to be unequal).  

\bigskip
$$\matrix{\epsfxsize=0.40\hsize\epsfbox{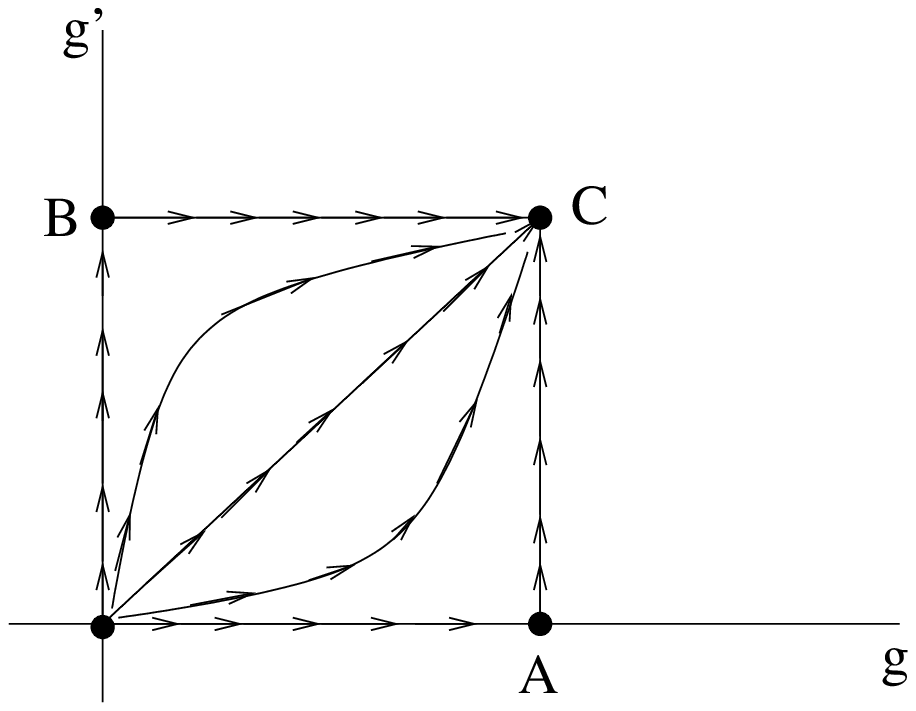} & \qquad & \epsfxsize=0.40\hsize\epsfbox{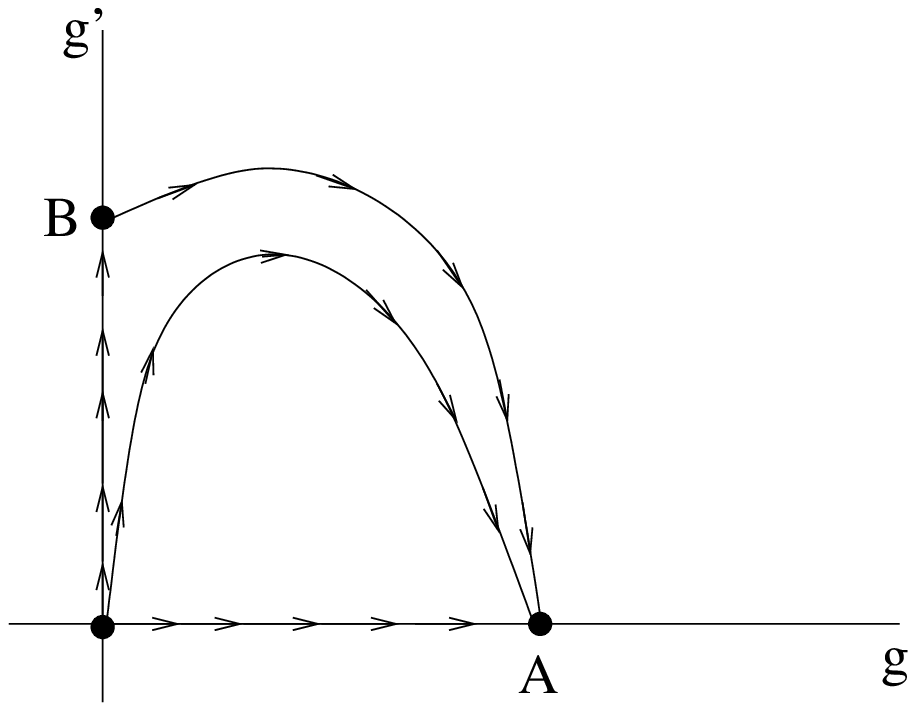} \cr
{\bf Figure\, 1:} & & {\bf Figure\, 2:} \cr
{\ninepoint \sl A\, and\, B\, are\, saddlepoints.} & & {\sl The\, plop.\, B\, is\, a\, saddlepoint.} \cr
{\sl C\, is\, stable,\, and\, there\, both} & & {\sl A\, is\, stable,\, and\, there\,  g'} \cr
{\sl groups\, are\, interacting}& & {\sl is\, driven\, to\, be\, IR\, free.}}$$

\bigskip
$$\matrix{\epsfxsize=0.40\hsize\epsfbox{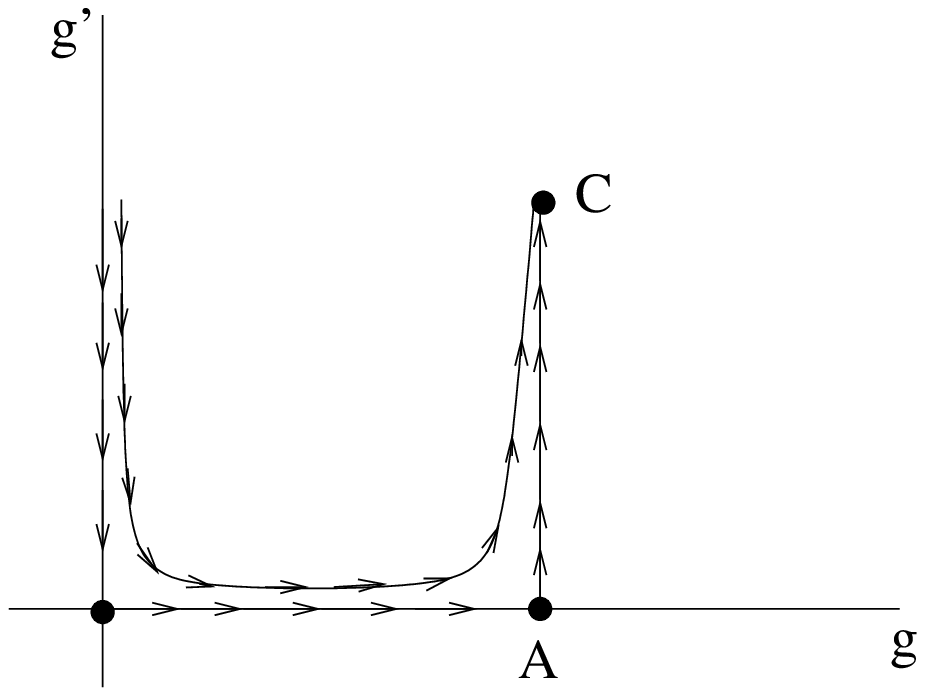} & \qquad & \epsfxsize=0.40\hsize\epsfbox{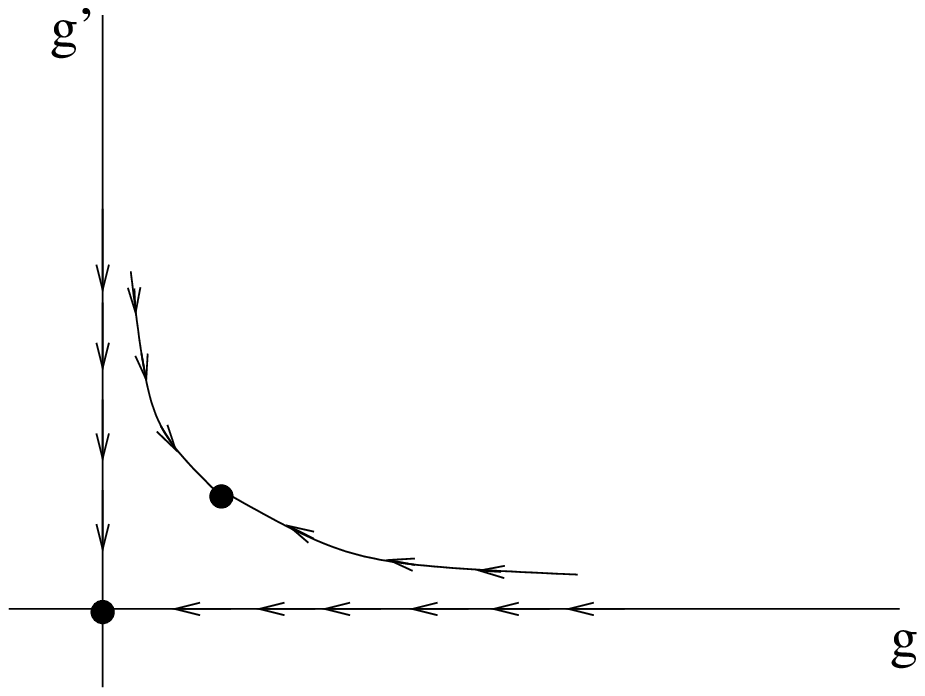} \cr
{\bf Figure\, 3:} & & {\bf Figure\, 4:} \cr
{\sl The\, opposite\, of\, fig.\, 2.} & & {\sl Two\, separately\, irrelevant\, couplings} \cr
{\sl g'\, is\, IR\, free\, for\, g=0\, but} & & {\sl combine\, to\, be\, interacting.} \cr
{\sl g \neq 0\, drives\, g'\, IR\, interacting.}& & {\sl \N=4\, SYM\, is\, such\, an\, example.}}$$

\bigskip

The theory \sunsun\ realizes the RG flows depicted in fig. 1 or fig. 2, depending
on the values of the parameters $(N_c, N_c', N_f, N_f')$.  With an additional superpotential,
as is present if we dualize one of the factors in \sunsun, the phenomenon of fig. 3 is also realized.

We'll focus here on supersymmetric theories, such as \sunsun\ with ${\cal N}=1$ supersymmetry,
where some exact results can be obtained.  However, we expect the phenomena of figs. 1 and 2 to occur
even in non-supersymmetric $G\times G'$ gauge theories, with matter in mixed $G\times G'$ representations, at least when the matter content is chosen such that each group is just barely asymptotically free.  There can then be RG fixed points in the perturbative
regime, as can be seen by considering the  beta functions to two
loops: 
\eqn\betabz{\beta _\alpha ={\alpha ^2\over 2\pi}(-b_1+b_2\alpha +c_2\alpha ')+O(\alpha ^4),\quad \hbox{and}\quad
\beta _{\alpha '}={\alpha '{}^2\over 2\pi}(-b_1'+b_2'\alpha '+c_2'\alpha )+O(\alpha ^4),}
(writing $\alpha = g^2/4\pi$ and $\alpha '=g'{}^2/4\pi$, and $O(\alpha ^4)$ refers to powers of either $\alpha$ or $\alpha'$), where the $c_2$ and $c_2'$ terms come from the matter in mixed representations; see e.g.
\twoloop.  Choosing the matter content to be such that the groups are barely asymptotically free, i.e. such that $b_1$ and $b_1'$ are small positive numbers, it is then found that the two-loop coefficients ($b_2$, $c_2$, $b_2'$, $c_2'$) in \betabz\ are positive and not especially small (e.g. in large $N_c$); this allows for RG fixed points to exist at relatively small
values of the fixed point coupling,  so that this argument for the RG fixed point's existence  could be qualitatively reliable.

In particular, to two loops, we find zeros of the beta functions \betabz\ at three points:
point (A) at $(\alpha _*, \alpha _*')_A\approx (b_1/b_2,0)$, point (B) at $(\alpha _*,\alpha '_*)_B \approx (0, b_1'/b_2')$, and point (C), at 
\eqn\bzfp{\pmatrix{\alpha _*\cr {\alpha '}_*}_C \approx {1\over b_2b_2'-c_2c_2'}\pmatrix{b_2'&-c_2\cr -c_2' &b_2}\pmatrix{b_1\cr b_1'}.}
For point (C) to actually exist, the values of  $\alpha _*$ and $\alpha '_*$ in \bzfp\ must be positive.  It is found 
that the determinant denominator in \bzfp\ is generally positive, so the positivity condition for RG fixed point
(C) to exist is thus
\eqn\bzcond{b_1b_2'>b_1'c_2\quad\hbox{and}\quad b_1'b_2>b_1c_2'
 \qquad\hbox{to have $g_*\neq 0$ and $g'_*\neq 0$.}}  These inequalities may or may not hold, depending on the choice of matter content.   The intuition for these inequalities is
that each gauge coupling makes the other less interacting in the IR (via the $c_2$ or
$c_2'$ terms), so there can only be a RG fixed point (C), with both interacting, if the couplings 
flow in balance: if either flows too much faster than the other, it can drive the other to be IR free.  For example, if the matter content is such that
$b_1c_2'>b_1'b_2$, then $g'\rightarrow 0$ in the IR,
as in fig. 2, with the $G$ dynamics overwhelming the $G'$ dynamics in the IR.  Likewise, if $b_1'c_2>b_1b_2'$, then $G'$ wins, and drives $g\rightarrow 0$ in the IR.  The inequality
$b_2b_2'>c_2c_2'$ implies that both inequalities in \bzcond\ could not be reversed.

The criteria \bzcond\ for RG fixed point (C) to exist are equivalent to the condition that RG fixed points (A) and (B) be IR unstable to perturbations in the other coupling, as depicted in fig. 5.  
For example, near (A), where $\alpha '=0$ and $\alpha _*\approx b_1/b_2$, \betabz\ gives
$\beta _{\alpha '}=\alpha '^2(-b_1'+c_2'b_1/b_2)/2\pi +O(\alpha '{}^3)$.  The second inequality in \bzcond\ is thus equivalent  to having (A) be IR repulsive to $\alpha'$ perturbations, as in fig. 1.
If  (A) and (B) are both IR repulsive to perturbations, generic couplings flow to having both 
interacting, and can end up at a fixed point
(C), as in fig. 1.   If either inequality of \bzcond\ is violated, then either (A) or (B) is IR attractive, and 
then  RG fixed point (C) does not exist
(at least it does not exist within the basin of attraction of zero couplings).  In that case, as depicted in 
fig. 2, generic RG flows attract to the IR stable point (A) or (B).  Because $b_2b_2'>c_2c_2'$, both
inequalities in \bzcond\ could not be reversed, i.e. we can not have  (A) and (B) both be 
IR attractive.  As depicted in fig. 6, such a hypothetical flow would have required an 
 unstable separatrix ridge, depicted as a dashed line, dividing the RG flows into two different
domains of attraction.  In neither the perturbative analysis, nor the supersymmetric examples to
follow, do we find examples of such flows. 

\bigskip
$$\matrix{\epsfxsize=0.40\hsize\epsfbox{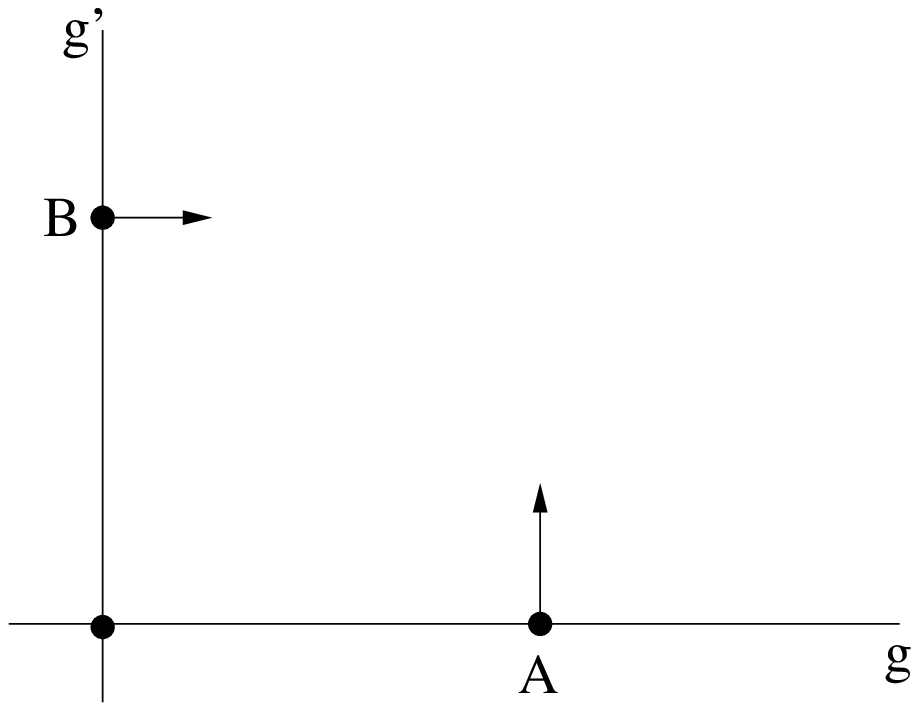} & \qquad & \epsfxsize=0.40\hsize\epsfbox{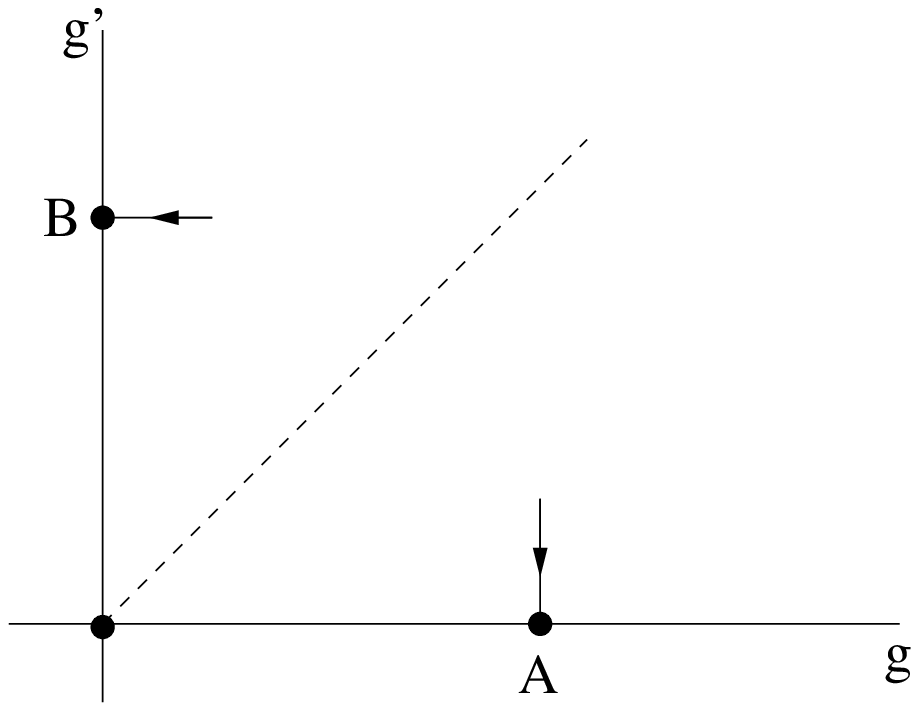} \cr
{\bf Figure\, 5:} & & {\bf Figure\, 6:} \cr
{\sl If\, A\, and\, B\, are\, both\, IR\, unstable\, to} & & {\sl We\, don't\, find\, examples\, of\, A\, and\, B\, both\, IR} \cr
{\sl perturbations,\, the\, theory\, flows\, to} & & {\sl stable\, to\, perturbations.\  Would've\, required\, a} \cr
{\sl C,\, with\, both\, couplings\, interacting.}& & {\sl separatrix\, between\, domains\, of\, attraction.}}$$
\bigskip

We can go beyond the above perturbative analysis in $\N =1$ supersymmetric theories, where exact results can be obtained via the NSVZ \NSVZ\ beta functions.  For a  
general $\N =1$ $G\times G'$ gauge theory, with
matter chiral superfield in representations $\oplus _i \bf{(r_i, r_i')}$, these are
\eqn\bnsvz{\eqalign{\beta _g(g, g')&=-{g^3f\over 16\pi ^2}
\left(3T(G)-\sum _i T(r_i)|r_i'|(1-\gamma _i(g,g'))\right)=-{3g^3f\over 16\pi ^2}\Tr ~ GGR\cr
\beta _{g'}(g, g')&=-{g'^3f'\over 16\pi ^2}
\left(3T(G')-\sum _i T(r_i')|r_i|(1-\gamma _i(g,g'))\right)=-{3g'^3f'\over 16\pi ^2}\Tr ~G'G'R.\cr}}
In the NSVZ scheme, $f=(1-{g^2T(G)\over 8\pi ^2})^{-1}$ and $f'=(1-{g'^2T(G')\over 8\pi ^2})^{-1}$, while
in other schemes these factors are replaced with other functions of the coupling \DKASlm, such that $f=1+O(g^2)$; these scheme-dependent prefactors are unimportant for our discussion, except for the 
fact that they should be strictly positive in our range of coupling constants. 

%\IntriligatorJJ
\nref\IW{
K.~Intriligator and B.~Wecht,
``The exact superconformal R-symmetry maximizes a,''
Nucl.\ Phys.\ B {\bf 667}, 183 (2003)
[arXiv:hep-th/0304128].
%%CITATION = HEP-TH 0304128;%%
}
\nref\AFGJ{D.~Anselmi, D.~Z.~Freedman, M.~T.~Grisaru and A.~A.~Johansen,
``Nonperturbative formulas for central functions of supersymmetric gauge
theories,''
Nucl.\ Phys.\ B {\bf 526}, 543 (1998)
[arXiv:hep-th/9708042].
%%CITATION = HEP-TH 9708042;%%
}
 \nref\AEFJ{D.~Anselmi, J.~Erlich, D.~Z.~Freedman and A.~A.~Johansen,
``Positivity constraints on anomalies in supersymmetric gauge
theories,''
Phys.\ Rev.\ D {\bf 57}, 7570 (1998)
[arXiv:hep-th/9711035].
%%CITATION = HEP-TH 9711035;%%
}
\nref\DKlm{
D.~Kutasov,
 ``New results on the 'a-theorem' in four dimensional supersymmetric field
theory,''
arXiv:hep-th/0312098.
%%CITATION = HEP-TH 0312098;%%
}
\nref\BIWW{
E.~Barnes, K.~Intriligator, B.~Wecht and J.~Wright,
``Evidence for the strongest version of the 4d a-theorem, via a-maximization
along RG flows,''
arXiv:hep-th/0408156.
%%CITATION = HEP-TH 0408156;%%
}
\nref\KPS{
D.~Kutasov, A.~Parnachev and D.~A.~Sahakyan,
``Central charges and U(1)R symmetries in N = 1 super Yang-Mills,''
JHEP {\bf 0311}, 013 (2003)
[arXiv:hep-th/0308071].}
%%CITATION = HEP-TH 0308071;%%
%\IntriligatorWR
\nref\IWbar{
K.~Intriligator and B.~Wecht,
``Baryon charges in 4D superconformal field theories and their AdS  duals,''
Commun.\ Math.\ Phys.\  {\bf 245}, 407 (2004)
[arXiv:hep-th/0305046].}
%%CITATION = HEP-TH 0305046;%%
\nref\Herzog{
C.~P.~Herzog and J.~Walcher,
``Dibaryons from exceptional collections,''
JHEP {\bf 0309}, 060 (2003)
[arXiv:hep-th/0306298].}
%%CITATION = HEP-TH 0306298;%%
%\IntriligatorMI
\nref\IWii{
K.~Intriligator and B.~Wecht,
``RG fixed points and flows in SQCD with adjoints,''
Nucl.\ Phys.\ B {\bf 677}, 223 (2004)
[arXiv:hep-th/0309201].}
%%CITATION = HEP-TH 0309201;%%
\nref\Franco{
S.~Franco, Y.~H.~He, C.~Herzog and J.~Walcher,
``Chaotic duality in string theory,''
arXiv:hep-th/0402120.}
%%CITATION = HEP-TH 0402120;%%
%\KutasovUX

\nref\CMT{
C.~Csaki, P.~Meade and J.~Terning,
``A mixed phase of SUSY gauge theories from a-maximization,''
JHEP {\bf 0404}, 040 (2004)
[arXiv:hep-th/0403062].
%%CITATION = HEP-TH 0403062;%%
}
%\BarnesJJ
%\BertoliniXF
\nref\BertoliniXF{
M.~Bertolini, F.~Bigazzi and A.~L.~Cotrone,
``New checks and subtleties for AdS/CFT and a-maximization,''
JHEP {\bf 0412}, 024 (2004)
[arXiv:hep-th/0411249].
%%CITATION = HEP-TH 0411249;%%
}
%\BenvenutiDY
\nref\DMJS{
S.~Benvenuti, S.~Franco, A.~Hanany, D.~Martelli and J.~Sparks,
``An infinite family of superconformal quiver gauge theories with
Sasaki-Einstein duals,''
arXiv:hep-th/0411264.
%%CITATION = HEP-TH 0411264;%%
}
The last equality in each line of \bnsvz\ involves  $\Tr~GGR$, which is the coefficient of the $U(1)_R$ ABJ triangle anomaly, with two external $G$ gluons.  This uses the fact that 
supersymmetry relates the dilatation current to a $U(1)_R$ current, with the exact scaling dimensions
of chiral fields related to their $U(1)_R$ charges:
\eqn\dimrc{\Delta ={3\over 2}R \quad\hbox{e.g.}\quad \Delta (Q_i)\equiv 1+\half \gamma _i={3\over 2}R(Q_i).}  When the theory is conformally invariant, this $U(1)_R$ is conserved, and part of the 
superconformal group $SU(2,2|1)$.  When the theory is not conformally invariant, e.g. along the RG flow {}from the UV to the IR, supersymmetry
still relates the stress tensor to an R-current, whose charges run with the anomalous dimensions
according to \dimrc, and whose anomaly is related to the beta function according to \bnsvz. 
Among all possible R-symmetries, the superconformal R-symmetry is that which locally maximizes
$a_{trial}(R)\equiv 3\Tr R^3-\Tr R$ \IW.  This method for determining the superconformal
R-charges is referred to as  ``a-maximization,"  because the value of $a_{trial}$ at its unique local maximum 
equals the conformal anomaly coefficient $a$ of the SCFT \refs{\AFGJ,  \AEFJ}\ (we rescale $a$ by a conventional factor of $3/32$).  An extension of a-maximization \DKlm, further explored in \refs{\BIWW,
\DKASlm}, has been proposed for determining the running R-charges, along the RG flow from the UV
to the IR.  See e.g. \refs{\KPS - \DMJS} for further applications and extensions of a-maximization.

For our particular example \sunsun, the exact beta functions \bnsvz\ are
\eqn\bnsvzsu{\eqalign{\beta _g(g, g')&=-{3g^3f\over 16\pi ^2}\Tr ~SU(N_c)^2R=-{g^3f\over 16\pi ^2}\left(b_1+N_f\gamma _Q+N_c'\gamma _X \right),\cr
\beta _{g'}(g, g')&=-{3g'^3f'\over 16\pi ^2}\Tr~SU(N_c')^2R=-{g'^3f'\over 16\pi ^2}\left(b_1' +N_f'\gamma _{Q'}+N_c\gamma _X\right),}}
where $b_1\equiv 3N_c-N_f-N_c'$ and $b_1'\equiv 3N_c'-N_f'-N_c$ are the one-loop beta functions.
We'll take both groups to be asymptotically free, i.e. take $g=g'=0$ to be UV attractive:
\eqn\susuaf{3N_c-N_f-N_c'>0, \quad \hbox{and}\quad 3N_c'-N_f'-N_c>0,}
so that $g=g'=0$ is IR repulsive, as in figs. 1 and 2.  To have the theory flow to a SCFT in the IR, rather than dynamically generating a vev, from a dynamically generated superpotential or quantum moduli space constraint, we also require
\eqn\susustab{N_f+N_c'>N_c\quad\hbox{and}\quad N_f'+N_c>N_c'\qquad\hbox{(stability)}.}

Assuming that \susuaf\ and \susustab\ 
hold, much as in the above perturbative analysis, we identify three possible RG fixed points:
\eqn\rgfpa{(A)\ g_*\neq 0,\ g'_*=0:\quad\hbox{where}\qquad  \gamma _Q=\gamma _X=-{b_1\over N_f+N_c'}, \quad \hbox{and}\quad \gamma _{Q'}=0.}
\eqn\rgfpb{(B)\ g_*=0, \ g'_*\neq 0: \quad\hbox{where}\qquad \gamma _{Q'}=\gamma _X=-{b_1'\over N_f'+N_c}, \quad \hbox{and}\quad \gamma _Q=0.}
\eqn\rgfpc{(C)\ g_*\neq 0,\ g'_*\neq 0: \quad \hbox{where}\quad b_1+N_f\gamma _Q+N_c'\gamma _X=0=  b_1'+N_f'\gamma _{Q'}+N_c\gamma _X.}
For point (A), we used the fact that there is an enhanced flavor symmetry which
ensures that $\gamma _X=\gamma _Q$ when $g'=0$, and that $Q'$ is a free field
for $g'=0$, so $\gamma _{Q'}=0$.  Analogous considerations apply for RG fixed point (B).  
Seiberg duality \NSd\ shows that (A) and (B) are actually interacting SCFTs only if 
\eqn\susunfm{N_f+N_c'>{3\over 2}N_c, \quad\hbox{and}\quad N_f'+N_c>{3\over 2}N'_c,}
respectively; otherwise (A) or (B) should be replaced with its free magnetic Seiberg dual.  

Our interest here is in the possible RG fixed point (C).  We'll discuss when it exists as an interacting SCFT.  We'll find, for example, that  \susunfm\ is modified, once the RG flow of both couplings is taken into account: the otherwise free magnetic dual can be driven interacting by the other gauge coupling, as depicted in fig. 3.  

Let us first discuss some simple necessary, though not sufficient, conditions for (C) to exist -- at
least within the domain of attraction of flows to the IR from the asymptotically free UV fixed point
at zero couplings -- by determining when the RG flow is as in fig. 1, or as in fig. 2, with one
of the couplings driven IR free.    (Our discussion here is 
essentially identical to one that already appeared in \PoppitzVH\ for a chiral example similar to \sunsun, having the field $X$ but not $\widetilde X$.)   As in fig. 5, (C) exists within the domain of attraction  of the UV fixed point only if (A) and (B) are both IR unstable to perturbations in the other coupling.  
Using \bnsvzsu, (A) is IR stable to $g'$
perturbations if $\Tr SU(N_f')^2R|_A<0$, i.e. we get $\beta _{g'}\sim -g'{}^3(b_1'-N_cb_1/(N _f+N_c'))+O(g'{}^5)$, with the second 
contribution from $\gamma _X$ at (A), so   $g'$ is an IR irrelevant perturbation of (A)  if $b_1'<N_cb_1/(N_f+N_c')$, i.e.
\eqn\rgfpas{\hbox{(A) is IR attractive, with $g'\rightarrow 0$, if $\qquad (3N_c'-N_f')(N_c'+N_f)-3N_c^2<0$.}}
Similarly, $g$ will be an irrelevant perturbation of (B) if $\Tr ~SU(N_c)^2R|_B<0$, which gives
\eqn\rgfpbs{\hbox{(B) is IR attractive, with $g\rightarrow 0$, if $\qquad (3N_c-N_f)(N_c+N_f')-3N_c'^2<0$.}}
The two inequalities in \rgfpas\ and \rgfpbs\ are mutually incompatible, so we do not find the
situation of fig. 6.  The condition for RG fixed point (C) to exist (within the domain of attraction of the UV fixed point) is that neither \rgfpas\ nor \rgfpbs\  holds, i.e. we have a flow as
in fig. 1 only if 
\eqn\bzexisint{(3N_c-N_f)(N_c+N_f')-3N_c'^2>0\qquad\hbox{and}\qquad
(3N_c'-N_f')(N_c'+N_f)-3N_c^2>0.}
The inequalities \bzexisint\ generally differ from the asymptotic freedom conditions \susuaf\ 
needed to have $g=g'=0$ not be IR attractive.  For the special case
$N_c=N_c'$ and $N_f=N_f'$, \bzexisint\ do reduce to the asymptotic freedom conditions \susuaf. 

When RG fixed point (C) does exist, the three independent anomalous dimensions, $\gamma _Q$, $\gamma _{Q'}$, and $\gamma _X$ are under-constrained by the two constraints of \rgfpc, so 
a-maximization \IW\ is required to  determine the exact anomalous dimensions of chiral operators at (C).  
When the RG fixed point is not at sufficiently strong coupling for there to be accidental symmetries, the 
a-maximization result can be written as \eqn\rxqq{\gamma _Q=1-\sqrt{1+{\lambda _{G}\over 2N_c}}, \quad \gamma _{Q'}=1-\sqrt{1+{\lambda_{G'}\over 2N_c'}}, \quad \gamma _X=1-\sqrt{1+{\lambda _{G}\over 2N_c}+
{\lambda _{G'}\over 2N_c'}},}
with $\lambda_{G}$ and $\lambda_{G'}$ determined by 
the two conditions in \rgfpc, for the two beta functions \bnsvzsu\ to vanish.  This way of writing 
the a-maximization result is motivated by the extension of a-maximization due to Kutasov \refs{\DKlm, \BIWW, \DKASlm},  where the interaction constraints on the 
superconformal R-charges, e.g. that the ABJ anomalies should vanish,  are imposed with Lagrange multipliers.  The conjecture is that the Lagrange multipliers can be interpreted as the running coupling constants along the flow to
the RG fixed point.  In particular, the claim is that \rxqq\ gives the running anomalous dimensions
along the entire RG flow, from $g=g'=0$ in the UV to the RG fixed point (C) in the IR, with $\lambda _G=g^2|G|/2\pi ^2$ and $\lambda _{G'}=g'{}^2|G'|/2\pi ^2$ the running couplings in some scheme.

This analysis needs to be supplemented when there are accidental symmetries \KPS, and we'll find that many accidental symmetries do arise in these theories for general
$(N_c, N_c', N_f, N_f')$.  a-maximization with many accidental symmetries is best left to a computer (we used Mathematica), and then it's simpler to do the a-maximization at the RG fixed point,
 imposing the constraints at the outset rather than with Lagrange multipliers.  We simplify the analysis by considering the limit of large numbers of 
flavors and colors, for arbitrary fixed values of the ratios, which for the example \sunsun\ are 
\eqn\ratios{x\equiv {N_c\over N_f}, \quad x'\equiv{N_c'\over N_f}, \quad n\equiv {N_f'\over N_f}.}
In this limit, the operator scaling dimensions then only depend on these ratios.  Depending on $(x,x', n)$, a variety of accidental symmetries associated with gauge invariant
operators hitting the unitarity bound are found to occur, and their effect on the a-maximization
analysis \KPS\ is accounted for in our numerical algorithm.

As a function of the parameters $(x,x',n)$, the theory either flows in the IR to a fully interacting RG fixed point, or
can be partially or fully free.   Our motivation for considering the examples \sunsun\ is that they have various possible dualities, and the a-maximization results can give insight into when they're applicable.  For example, we could
Seiberg dualize \NSd\ one of the groups in \sunsun, treating the other as a weakly gauged spectator.
As we'll discuss, there is a range of $(x,x',n)$ for which this dual theory realizes the RG flow possibility
of fig. 3: an arbitrarily small non-zero coupling of the ``spectator"  group can drive an otherwise free magnetic group to be interacting in the IR.    This
also occurs in an example discussed in \CMT, which appeared during the course of the present work.

Knowing the exact dimensions of chiral operators, we can classify the relevant superpotential deformations of (C).  
In particular, we can now determine the ``superconformal window" range of validity of a duality
proposed in \ILS\ for the theory \sunsun\ with added superpotential interaction  
$W_{A_{2k+1}}=\Tr (X\widetilde X)^{k+1}$.  The dual \ILS\ has 
gauge group 
$SU((k+1)(N_f+N_f')-N_f-N_c')\times SU((k+1)(N_f+N_f')-N_f'-N_c)$ with similar matter content
and additional gauge singlets
(corresponding to the mesons), and a dual analog of the $W_{A_{2k+1}}$ superpotential.   The
superconformal window, where both dual descriptions are useful, is the range of 
$(N_c, N_c', N_f, N_f')$ within which both the electric $W_{A_{2k+1}}$, as well as its analog in the magnetic dual, are both relevant.  The a-maximization results allow us to determine this 
subspace of $(x,x',n)$ parameter space, as a function of $k$.   For large $k$, we find that this subspace is necessarily close to
the $x\approx x'$ slice, i.e. $N_c\approx N_c'$.

The outline of this paper is as follows. In sect. 2 we briefly review a-maximization, and apply it to 
determine  the superconformal R-charges for the $SU(N_c)\times SU(N_c')$ example \sunsun.
 We find that there are accidental symmetries arising from gauge invariant operators hitting
the unitarity bound $\Delta \geq 1$, and use the procedure of \KPS\ to take these into account during a-maximization.  We especially consider the parameter slice $x=x'$ (i.e.  $N_c = N_c'$) for large $x$ (i.e. 
$N_c\gg N_f$), and general $n$.   In this slice and limit, $R(X)\rightarrow 0$.  We account for the many accidental  symmetries, associated with generalized mesons hitting their unitarity bound, in this limit (and numerically check that no baryon operators hit the unitarity bound.)    As we discuss, if we set
$n\equiv N_f'/N_f=1$, our results should -- and indeed do -- coincide with those of \KPS.

In sect. 3 we consider the theory \sunsun\ deformed by the superpotential $W_{A_{2k+1}}=\Tr (X\widetilde X)^{k+1}$, and the dual description of \ILS\ of that theory.  We use a-maximization
to determine the exact chiral operator dimensions in the dual of \ILS.  Combining these results
with those of sect. 2, we can  determine the superconformal window region of $(x,x',n)$ parameter space, for any given value of $k$, 
within which the $W_{A_{2k+1}}$ superpotential of both the electric theory \sunsun\
and its dual are both relevant.   For large $k$, the superconformal window is necessarily close to the 
parameter slice $x\approx x'$.  We check numerically that, for all $k$,  there
is always a non-empty superconformal window region of parameter space in which the duality of \ILS\ 
is applicable.  

In sect. 4 we consider Seiberg dualizing \NSd\ one of the groups in \sunsun, treating the other gauge
group as a spectator.   We'll discuss analogs $\widetilde A$, $\widetilde B$, and $\widetilde C$
of the possible RG fixed points in fig. 1, when they exist, and when they're IR stable to perturbations.
We find that there is a range of the parameters \ratios\ $(x,x',n)$ where an otherwise IR free magnetic
gauge group is driven to be interacting for any non-zero gauge coupling of the ``spectator" group.
This is the phenomenon depicted in fig 3.    As seen from the exact beta functions \bnsvz, positive
anomalous dimensions are needed to turn a 1-loop IR irrelevant coupling into an IR relevant one. 
The superpotential of the Seiberg dual theory 
plays a crucial role here, together with the spectator gauge coupling, to get the positive anomalous
dimensions needed for the effect of fig. 3.   The condition that the RG fixed point $(C)$ of the 
original electric theory \sunsun\ be interacting, rather than flowing to a free magnetic dual, 
is that the fully interacting RG fixed point $(\widetilde C)$ exists in the dual theory; this issue is analyzed by the dual analog of fig. 5.  When
$(\widetilde C)$ does exist, we expect that it's equivalent to the
RG fixed point $(C)$ of the electric description.  We verify that their superconformal R-charges and
central charges indeed agree.  

In Sect. 5 we briefly conclude, and present a topic for further research. 

In the Appendix, we note that all of the many duality examples of \ILS\ have a non-zero superconformal 
window.   The theories in \ILS\ with a single gauge group (either $SU(N_c), SO(N_c),$ or $Sp(N_c)$)
and matter in a two-index representation (e.g. adjoint, symmetric, or
antisymmetric) are shown in the large $N_c$ limit to all have the same superconformal R-charges,  and superconformal window, as that of $SU(N_c)$ with an adjoint; we can directly borrow the
results obtained in \KPS, with the central charge differing from that of \KPS\ by just a fixed
overall multiplicative factor.  
We also note that all of the other examples in \ILS, involving product groups, all also have superconformal R-charges and superconformal window that reduce to those obtained
in \KPS\ for a 1d slice of their parameter space, when we take all of the group ranks equal
and all numbers of flavors equal  (e.g. taking $x=x'$ and $n=1$ in \ratios).  This suffices to
show that all of the duality examples of \ILS\ indeed have a non-empty superconformal window.

\newsec{ a-maximization analysis for the $SU(N_c)\times SU(N_c')$ theory \sunsun}

The superconformal $U(1)_R$ symmetry is uniquely determined by the condition that
it maximizes $a_{trial}(R)=3\Tr R^3-\Tr R$ among all possible R-symmetries \IW.  The constraints
on the superconformal R-symmetry, e.g. that it's ABJ anomaly free, can either be implemented
at the outset, before maximizing $a(R)$ w.r.t. $R$, or via Lagrange multipliers $\lambda$ \DKlm.  a-maximization with the Lagrange multipliers yields simple general expressions for the R-charges
of the fields $R_i(\lambda)$, with the conjectured interpretation of giving the running R-charges
along the RG flow to the RG fixed point \refs{\DKlm, \BIWW, \DKASlm}.  

For example, for a general ${\cal N}=1$ supersymmetric 
 $G\times G'$ gauge theory, with zero superpotential, we determine the running R-charges by
maximizing  with respect to the $R_i$
\eqn\ggalm{\eqalign{a(\lambda, R)&=3\Tr R^3-\Tr R-\lambda _G\Tr \, G^2R-\lambda _{G'}\Tr \, G'{}^2R
=2|G|+2|G'|-\lambda _GT(G)-\lambda _{G'}T(G')\cr &+\sum_i|{\bf r_i}||{\bf r'_i}|\left[
3(R_i-1)^2-1-
\lambda _G{T({\bf r_i})\over |{\bf r_i}|}-\lambda _{G'}{T({\bf r_i'})\over |{\bf r_i'}|}\right](R_i-1),}}
holding fixed the Lagrange multipliers $\lambda _G$ and $\lambda _{G'}$,
which enforce the constraints that $U(1)_R$ not have ABJ anomalies, $\Tr GGR=\Tr G'G'R=0$.
This yields:
\eqn\rfprod{R_i(\lambda) =1-{1\over 3}\sqrt{1+\lambda _G{T ({\bf r}_i)\over |{\bf r}_i|}+\lambda _{G'}
{T  ({\bf r'}_i)\over |{\bf r'}_i|}}\quad\hbox{i.e.}\quad \gamma _i=1-\sqrt{1+\lambda _G{T ({\bf r}_i)\over |{\bf r}_i|}+\lambda _{G'}
{T  ({\bf r'}_i)\over |{\bf r'}_i|}}}
where we used $\gamma_i = 3R_i-2$ for the anomalous dimensions. The conjecture is that these
expressions can be interpreted as giving the anomalous dimensions along the entire RG flow, with 
$\lambda _G=g^2|G|/2\pi ^2$ and $\lambda _{G'}=g'{}^2|G'|/2\pi ^2$ in some scheme.
For the example \sunsun\
this gives the result \rxqq.
As in \DKlm, using \rfprod\ in 
\ggalm\ yields a monotonically decreasing a-function $a(\lambda)=a(\lambda, R(\lambda))$ along the
RG flow.  The values of $\lambda ^*_G$  and $\lambda ^*_{G'}$ at the IR fixed point are the extremal values of $a(\lambda)$; since $a(\lambda)$'s gradients are proportional to the exact beta functions 
\refs{\DKlm, \BIWW, \DKASlm}, this is equivalent to the conditions that the anomalous dimensions \rxqq\ yield zeros of the beta functions \bnsvz.

Whenever a gauge invariant operator $M$ hits or appears to violate the unitarity bound $R(M)\geq 2/3$, $M$ becomes
a decoupled free field.  This affects the a-maximization analysis by introducing an additive
correction to the quantity $a(R)$ to be maximized \KPS\ (this can be derived from the presence of
an accidental $U(1)_M$ symmetry, acting only on $M$ \IWii):
\eqn\acorr{a_{trial}(R)\rightarrow a_{trial}(R)+{1\over 9}dim(M)\left(2-3R(M)\right)^2\left(5-3R(M)\right).}
This correction can also be included in the a-maximization analysis with Lagrange multipliers
\BIWW, but it becomes unwieldy to do so when there are many such contributions from operators
that hit the unitarity bound, as is the case in our examples for general values of the numbers of flavors and colors.  For this reason, we will here do the a-maximization analysis at the RG fixed point, numerically, with the constraints implemented at the outset rather than via Lagrange multipliers.

We consider the example \sunsun\ in the range of the parameters \ratios\ where it's possible to have the RG fixed point like (C) in fig. 1, with both groups interacting.  For asymptotic freedom of
$g=g'=0$ in the UV, and to avoid having it be attractive in the IR, we take 
\eqn\susuasympf{3x-x'-1>0, \quad\hbox{and}\quad 3x'-x-n>0. }
We also impose the condition \susustab, which is
\eqn\susustabx{-n<x-x'<1 \qquad \hbox{(stability)},}
to have the origin of the moduli space of vacua not be dynamically disallowed.  Finally, to have
the points (A) and (B) not be IR attractive, as in fig. 2, we impose
\bzexisint, 
\eqn\susuintir{(3x-1)(x+n)-3x'{}^2>0 \quad\hbox{and}\quad(3x'-n)(x'+1)-3x^2>0.}  If
either inequality of \susuintir\ is not satisfied, one or the other group is driven IR
free, to RG fixed point (A) or (B), with anomalous dimensions and R-charges given by
\rgfpa\ or \rgfpb.  When both \susuintir\ are satisfied, RG flows generally end up with both
couplings interacting, which can end up being a 
RG fixed point (C), where \rgfpc\ is satisfied.  As mentioned in the introduction, we do {\it not}
impose the naive conditions \susunfm\ to avoid IR free magnetic dual groups: 
as we'll see in sect. 4, the conditions \susunfm\ are generally 
dramatically modified by the dynamics of the other gauge group.

As always, the conditions in \rgfpc\ for the exact beta functions to vanish are equivalent to requiring that the superconformal $U(1)_R$ have vanishing ABJ anomalies 
 with respect to all of the interacting gauge groups.  So at
(C) we have the two anomaly free conditions 
\eqn\sunsunaf{\eqalign{N_c+N_c'(R(X)-1)+N_f(R(Q)-1)&=0\cr
N_c'+N_c(R(X)-1)+N_f'(R(Q')-1)&=0}}  
to have $\Tr SU(N_c)^2U(1)_R=\Tr SU(N_c')^2U(1)_R=0$.  
Enforcing \sunsunaf\ at the RG fixed point, we can solve for $R(X)$ and $R(Q')$ in terms
of  $y\equiv R(Q)$
\eqn\rxyp{
    R(X) = {1-y\over x'} + 1-{x\over x'}, \qquad
    R(Q') = {x\over nx'}(y-1) + {x^2\over nx'} - {x'\over n} +
    1,}
where the parameters $(x,x',n)$ of the theory are the ratios \ratios.  We  determine
the superconformal R-charge $y(x,x',n)$ by a-maximization (in the single variable $y$).

Imposing \rxyp, we compute $a_{trial}(R)=3\Tr R^3-\Tr R$ from the spectrum \sunsun\  to be
\eqn\aosusu{\eqalign{a
^{(0)}/N_f^2 &= 2x^2+2x'^2+6x(y-1)^3 - 2x(y-1) + 6nx'\left [{x\over nx'}(y-1)
+ {x^2\over nx'} - {x'\over n}\right]^3  \cr &- 2nx'\left
[{x\over nx'}(y-1) + {x^2\over nx'} - {x'\over n}\right ] +
6xx'\left [{1-y\over x'} -{x\over x'}\right ]^3 - 2xx'\left
[{1-y\over x'} -{x\over x'}\right ].}}
We then compute the superconformal R-charges by locally maximizing \aosusu\ w.r.t. $y$, for general fixed values of the parameters $(x,x',n)$; we'll denote the solution as $y^{(0)}(x,x',n)$.    The superscript ${}^{(0)}$ is a reminder that these results are valid only in the range of $(x,x',n)$
in which no gauge invariant operators have hit the unitarity bound; otherwise \aosusu\ will
require corrections as in \acorr.  Within this range of $(x,x',n)$,  we can also use the Lagrange multiplier approach. Imposing \sunsunaf\ with Lagrange multipliers yields
the simple expressions \rxqq, which can be interpreted as the running R-charges along the RG flow, coinciding with the R-charges obtained above from $y^{(0)}(x,x',n)$ at the RG fixed point.  

The first gauge invariant, chiral, composite operators ${\cal O}$ to hit the unitarity bound $R({\cal O})\geq 2/3$ are the mesons $M\equiv Q\widetilde Q$ or $M'\equiv Q'\widetilde Q'$.  $M$ hits the unitarity
bound when $y(x,x',n)\leq 1/3$; using \dimrc, this happens when $Q$ has the large, negative
anomalous dimension, $\gamma _Q(x,x',n)<-1$, which can only happen if $(x,x',n)$ are such that the RG fixed point values of the gauge couplings are large.  The above result $y^{(0)}(x,x',n)$ is valid within
the range of $(x,x',n)$ where neither $M$ or $M'$ have hit their unitarity bound, i.e. the range of 
$(x,x',n)$ where $y^{(0)}(x,x',n)> 1/3$ and where 
$R(Q')\geq 1/3$,  with $R(Q')$ computed from $y^{(0)}(x,x',n)$ via \rxyp.    Outside of this range, the
above a-maximization analysis has to be supplemented, as in \acorr,  to account for the accidental symmetries associated with operators hitting the unitarity bound and becoming free fields. 
  
For general $(x,x',n)$ the gauge operators that will hit the unitarity bound are:
\eqn\susumesons{M_j= Q(\widetilde XX)^{j-1}\widetilde Q, \ M_j' = Q'(\widetilde XX)^{j-1}\widetilde Q', \  P_j = Q(\widetilde XX)^{j-1}\widetilde X\widetilde Q', \  {\widetilde P}_j = \widetilde QX(\widetilde
    XX)^{j-1}Q'.}  For every integer $j\geq 1$, there are
$N_fN_f'$ mesons $P_j$ and $\widetilde P_j$, $N_f^2$ mesons $M_j$, and $N_f'{}^2$ mesons
$M'_j$.  We verified that it's self-consistent to assume that the baryon operators do not hit the unitarity bound; also, gauge invariants without fundamentals, such as $\Tr (X\widetilde X)^j$, contribute negligibly in the large $N$ limit.   The quantity to maximize in general is then
\eqn\eqnnine{\eqalign{
    a ^{(p)}/N_f^2 &=\at ^{(0)}/N_f^2 +
    {2n\over 9}\sum_{j=1} ^{p_P} \left [2-3R(P_j)\right ]^2\left [5-3R(P_j)\right ]
    \cr &+
    {1\over 9}\sum
    _{j=1} ^{p_M} \left [2-3R(M_j)\right ]^2\left [5-3R(M_j)\right ] + {n^2\over 9}\sum
    _{j=1} ^{p_{M'}} \left [2-3R(M_j')\right ]^2\left
    [5-3R(M_j')\right],}}
where $p$ denotes $\{p_P,p_M,p_{M'}\}$, with $p_{P}=p_{\widetilde P}$ the number of $P$
(and also $\widetilde P$ type) mesons which have hit the unitarity bound.   The quantities
such as $R(M_j)$ in \eqnnine\ are given by e.g. $R(M_j)=R[Q(X\widetilde X)^{j-1}\widetilde Q]=2y+2(j-1)R(X)$, with $R(X)$ given by \rxyp; so the corrections in \eqnnine\ are complicated
functions of the variable $y$ that we're maximizing with respect to, along with the parameters
$(x,x',n)$.  Maximizing \eqnnine\ yields
$y^{(p)}(x, x', n)$, and $y(x,x', n)$ is obtained by pasting these together, with the appropriate values
of $p$ depending on $(x,x',n)$, increasing e.g. $p_M$ every time another value of $j$ is obtained
such that $R(M_j)$  hits $2/3$.    We numerically implemented this process to obtain $y(x,x',n)$, but
it's difficult to produce an illuminating plot of a function of three variables.  

Let us discuss some qualitative aspects of our results.  From $y(x,x',n)$ we can compute
the anomalous dimensions $\gamma _Q$, $\gamma _X$, and $\gamma _{Q'}$, using \dimrc,
and we find that all are negative within the range \susuintir\ where the RG fixed point (C) can
exist.  This is to be expected, since our theory \sunsun\ has only gauge interactions, and no
superpotential (gauge interactions yield negative anomalous dimension, and superpotentials
yield positive contributions to the anomalous dimensions).   When either inequality \susuintir\ is violated, the theory flows not to (C), but rather to RG fixed points (A) or (B), as in fig. 2 and the above a-maximization analysis, which assumed in 
\sunsunaf\ that both groups are interacting, is inapplicable.  At the boundaries of \susuintir,
where either inequality is saturated, our a-maximization results properly reduce to \rgfpa\ or
\rgfpb. 

It is interesting to note that there is a 1d slice of the $(x,x',n)$ parameter space, given by $x=x'$ and $n=1$, for which the a-maximization analysis of this theory coincides with that of \KPS\ for $SU(N_c)$ SQCD with $N_f$ fundamentals and an added adjoint. In this slice, for every contribution to the
quantity $a_{trial}$ to maximize in \KPS, we have here two analogous matter fields in the spectrum 
of our theory, with the same R-charges: twice as many gauge
fields, twice as many fundamentals ($Q$ and $Q'$ and conjugate), the $X$ and $\widetilde X$ fields
contribute as two adjoints (using \rxyp\ for $x=x'$ and $n=1$), and all the mesons \susumesons\ hitting the unitarity bound map to two
copies of the mesons hitting the unitarity bound in \KPS.  Thus, for $x=x'$ and $n=1$, \eqnnine\  
is exactly twice the expression obtained in \KPS\ for the theory considered there.  Since $a_{trial}$ is the
same function of $y$, up to a factor of 2, it is maximized by the same solution $y_{KPS}(x)$ obtained by the analysis of \KPS.  So the 
superconformal R-charges of our theory \sunsun\ satisfy $y(x,x',n)|_{x=x', n=1}=y_{KPS}(x)$.

For $x\approx x'$ taken to be very large, i.e. $N_c\approx N_c'\gg N_f$, the superconformal R-charge of the field $X$ goes to zero, $R(X)\rightarrow 0$, for arbitrary fixed values of $n\equiv N_f'/N_f$, as seen from \rxyp, and the fact that $y$ remains finite in this limit. The asymptotic value $y_{as}(n)$ in
this $x=x'\rightarrow \infty$ limit is determined by our numerical a-maximization analysis, but it can 
also be approximated analytically. Because many mesons
contribute to the sums \eqnnine, the sums can be approximated as integrals (following \KPS):
\eqn\mesonsa{{1\over 9}\sum _{j=1}^{p}[2-3R_j]^2[5-3R_j]
\approx
{1\over 27\beta}\int _0^{2-3\alpha}u^2(3+u)du={1\over 18 \beta}(2-3\alpha)^3(1-\half \alpha)}
where $\alpha$ and $\beta$ are defined by $R_j\equiv \alpha +(j-1)\beta$ (and $u\equiv 2-3R_j$).
The upper limit $p$ in the sum is solved for by setting $R_p=\alpha +(p-1)\beta$ equal to $2/3$.
Applying \mesonsa\ to the sums in \eqnnine, $\beta= R(X\widetilde X)=2R(X)$ for all three, and for the first sum in \eqnnine\  $\alpha = R(Q\widetilde X \widetilde Q')=y+R(Q')+R(X)$, while for the second and third $\alpha = 2y$ and $\alpha = 2R(Q')$ respectively; here,  $R(X)$ and $R(Q')$ are to be written in terms of the variable $y$ and the parameters $(x,x',n)$
using \rxyp.  

Setting $x=x'$ and taking both large, \eqnnine\ then becomes
\eqn\eqnthirteen{\eqalign{    a/N_f^2 &\simeq 6x\left [1+{1\over n^2}\right ](y-1)^3 -
    20x(y-1) \cr &+ {x\over 36}\left [2-6y\right ]^3 + {xn\over 36}\left [{6\over
    n}(1-y)-4\right]^3+{xn\over 36}\left (1+{1\over n}\right )\left [3(1-y)\left (1+{1\over n}\right )-4\right]^3.}}
The first line of \eqnthirteen\ is the large $x=x'$ limit of \aosusu, and the second line contains the meson
sums of \eqnnine, evaluated using \mesonsa.  Note that every term in \eqnthirteen\ is linear\foot{The fact that the
expression in \eqnthirteen\ grows for large $x$ only linearly is a 
check of the conjectured a-theorem.  Any greater exponent would've led to a-theorem violations, e.g. along a Higgs flat direction where $X\widetilde X$ gets an expectation value, Higgsing
each $SU(N_c)$ gauge group factor to products of similar factors.  This flat direction is analogous to 
that considered in a non-trivial check of the
a-theorem in \KPS\ (where it's also pointed out that the sub-leading constant term must be -- and indeed is --
negative for the a-theorem to hold for this Higgs RG flow).}   in $x$ in 
this limit, so maximizing \eqnthirteen\ w.r.t. $y$ yields an 
asymptotic value, $y_{as}(n)$, that's independent of $x$ in this limit of large $x=x'$.  
   This asymptotic value depends
on $n\equiv N_f'/N_f$, and the conditions \susuintir\ needed for neither gauge coupling to drive the other to be IR free here require $n$ to lie in the range
\eqn\nrange{3>n>{1\over 3}\qquad\hbox{for $x=x'\rightarrow \infty$}.}
As expected from the discussion above, for $n=1$ \eqnthirteen\ reduces to twice 
the expression obtained in the large $x$ analysis of \KPS, and for $n=1$ our expression for 
$y_{as}(n)$ coincides with the asymptotic large $x$ value of $y$ obtained there: $y_{as}(n)|_{n=1}=
(\sqrt{3}-1)/3$.

The asymptotic value $y_{as}(n)$ will be used in the next section to find the minimal value of 
$x\approx x'$ needed for the superpotential $\Delta W _{A_{2k+1}}\equiv  \Tr (X\widetilde X)^{k+1}$ to be a relevant deformation of RG fixed point (C) in the limit of large $k$.  This gives one side
of the superconformal window for the duality of \ILS\  (see fig. 7).  We have also checked, including away from the strict  $x=x'$ limit,  that $R(X)$ is nowhere 
negative, i.e. using \rxyp\ that the  a-maximizing solution $y(x,x',n)$ satisfies 
$1-y(x,x',n)+x'-x>0.$  This is 
important for the self-consistency of our analysis  since, if $R(X)$ were negative,
 baryonic operators, formed by dressing the quarks with many powers of $X\widetilde X$, would 
 hit the unitarity bound and lead to additional contributions analogous to \acorr.

\newsec{The theory with $W_{A_{2k+1}}=\Tr (X\widetilde X)^{k+1}$ and its dual}

In \ILS\ it was proposed that our theory \sunsun, together with a superpotential $W_{A_{2k+1}}=\Tr (X\widetilde X)^{k+1}$ has a dual given by a similar theory: 
\eqn\sunsund{\matrix{&&\hbox{gauge group:} &SU(\widetilde N_c)\times SU(\widetilde{N}_c')&\cr
&\hbox{matter:} &Y\oplus \widetilde Y&\bf{(\fund , \overline{\fund} )}\oplus \bf{(\overline{\fund},
\fund )},&\cr
&&q_f \oplus \widetilde q_{\tilde f}&\bf{(\fund , 1)} \oplus \bf{(\overline{\fund}, 1)}&(f, \widetilde f=1\dots N_f'),\cr
&&q'_{f'}\oplus \widetilde q'_{\tilde f'}& \bf{(1, \fund )}\oplus \bf{(1, \overline{\fund})} &(f',\widetilde f'=1\dots N_f), \cr}}
where
$\widetilde{N}_c=(k+1)(N_f+N'_f)-N_f-N'_c$ and $\widetilde{N}'_c=(k+1)(N_f+N'_f)-N'_f-N_c$.  There are also singlets $P_j$, for $j=1\dots k$, and $M_j$ and $M'_j$ for $j=1\dots k+1$, with superpotential
 \eqn\eqnfourteen{\eqalign{
    W=\Tr (Y\widetilde Y)^{k+1}&+\sum ^k _{j=1}\left[ P_j q\widetilde Y(Y\widetilde Y)^{k-j}\widetilde q'+{\widetilde P}_j \tilde q'(Y\widetilde Y)^{k-j}Yq \right] \cr &+\sum
    ^{k+1} _{j=1} \left [ M_j q'(\widetilde YY)^{k-j+1}\tilde q' +
    M'_j\tilde q(Y\widetilde Y)^{k-j+1}q\right].}}
 The first term is the dual analog of the $W_{A_{2k+1}}$ superpotential, and the remaining terms
 are analogs of the $Mq\widetilde q$ superpotential in Seiberg duality \NSd.
 
The duality is useful within a superconformal window, which is the range of the parameters $(x,x',n)$, where the superpotential $W_{A_{2k+1}}$ and its dual analog in \eqnfourteen\ are both relevant in controlling the IR dynamics, i.e. when the following conditions are satisfied:
\eqn\welrel{(i)\qquad R(X)={1-y(x,x',n)\over x'}+1-{x\over x'}<{1\over k+1},}
\eqn\wmagrel{(ii)\qquad R(Y)={1-\widetilde y(\widetilde x, \widetilde x', \widetilde n)\over \widetilde x'}+1-{\widetilde x \over \widetilde x'}<{1\over k+1},}
\eqn\stab{(iii)\qquad (k+1)(N_f+N_f')-N_f-N_c'>0, \quad\hbox{and}\quad (k+1)(N_f+N_f')-N_f'-N_c>0.}
If $(i)$ is not satisfied, $W_{elec}=\Tr (X\widetilde X)^{k+1}$ is an irrelevant deformation of RG fixed point (C), and thus
$W_{elec}\rightarrow 0$ in the IR; this fact is  obscured in the magnetic dual description.
Likewise, if $(ii)$ is not satisfied, one should use the magnetic description, with the coefficient of the $\Tr (Y\widetilde Y)^{k+1}$ superpotential term flowing to zero in the IR; the electric description then
doesn't readily describe the true RG fixed point.  Finally, condition $(iii)$ is the ``stability bound," 
needed for the RG fixed point to exist (and for the dual groups \sunsund\ to have positive ranks): if \stab\ are  not satisfied, the electric theory \sunsun\ with $W_{A_{2k+1}}$ 
superpotential dynamically generates a superpotential, spoiling conformal invariance.  

Using the results of the previous subsection, we can now determine the 
range of $(x,x',n)$ in which condition \welrel\ is satisfied, for $\Tr (X\widetilde X)^{k+1}$ to be relevant.  
The $k=0$ case is a mass term and \welrel\ is then always satisfied (starting with $W=0$ at (C), all fields
have $R\leq 2/3$).
For all $k>1$, \welrel\ is only satisfied 
in subspaces of the $(x,x',n)$ parameter space for which the 
RG fixed point is at sufficiently strong enough coupling to give $X$ a sufficiently negative anomalous dimension.  The larger $k$ is, the more strongly coupled the RG fixed point must be in order to have
\welrel\ be satisfied.   For arbitrarily large $k$, there's a non-empty range of $(x,x',n)$ in which \welrel\
is satisfied: as we saw in the previous subsection, $R(X)\rightarrow 0$ in parts of the parameter space.
Let's consider, for example, the parameter slice $x=x'$ and ask when \welrel\ is satisfied for large values of $k$.  Since satisfying \welrel\ for large $k$ requires large $x$, we can replace $y(x,x',n)$ in \welrel\ 
with the asymptotic value $y_{as}(n)$ obtained by maximizing \eqnthirteen.  Then the condition \welrel\ for the superpotential to be relevant becomes 
\eqn\xminn{x>x_{min}(n)\approx k\left( 1-y_{as}(n)\right) \qquad\hbox{for $k\gg 1$}.}

The above analysis of the electric theory gives one edge of the superconformal window
of the parameters $(x,x', n)$ for the duality of \ILS.  The other edge of the window is obtained
by determining the range of these parameters in which \wmagrel\ is satisfied, for the $W_{A_{2k+1}}$
superpotential to be relevant in the magnetic theory.   Again, we simplify the analysis by taking the numbers of flavors
and colors in the electric theory to be large, so the same is true in the magnetic theory. 
The ratios on the magnetic side are
defined to be $\xt\equiv\widetilde{N}_c/N'_f$, $\xt
'\equiv\widetilde{N}'_c/N'_f$, and $\nt\equiv N_f/N'_f$, which are related to the electric ones \ratios\
as
\eqn\emratior{\xt = (k+1)(1+n^{-1})-n^{-1}-x'n^{-1}, \quad \xt ' = (k+1)(1+n^{-1})-1-xn^{-1}, \quad
\nt = n^{-1}.}
In the magnetic theory \sunsund, we 
assume that both magnetic gauge groups remain interacting. The superconformal R-charge is
constrained by the magnetic analog of \sunsunaf, that it be anomaly free w.r.t. both gauge groups.
As in \rxyp, we can use this to solve for $R(Y)=R(\widetilde Y)$ and $R(q')=R(\widetilde q')\equiv \widetilde y'$ in terms of $R(q)=R(\widetilde q)\equiv \widetilde y$:
\eqn\rxypd{R(Y) = {1-\widetilde y\over \widetilde x'} + 1-{\widetilde x\over \widetilde x'}, \qquad
    \widetilde y' = {\widetilde x\over \widetilde n\widetilde x'}(\widetilde y-1) + {\widetilde x^2\over \widetilde n\widetilde x'} - {\widetilde x'\over \widetilde n} +
    1.}

The contribution to the magnetic $\widetilde a_{trial}$ from the fields in \sunsund\ is
\eqn\eqnsixteen{\eqalign{
    \at ^{(0)}/N_f'^2 &= 2\xt^2+2\xt'^2+6\xt(\yt-1)^3 - 2\xt(\yt-1) +
    6\nt\xt'\left [{\xt\over\nt\xt'}(\yt-1) + {\xt^2\over\nt\xt'} - {\xt'\over\nt}\right]^3
    \cr &-
    2\nt\xt'\left [{\xt\over\nt\xt'}(\yt-1) + {\xt^2\over\nt\xt'} - {\xt'\over\nt}\right ] +
    6\xt\xt'\left [{1-\yt\over\xt'} -{\xt\over\xt'}\right ]^3 - 2\xt\xt'\left [{1-\yt\over\xt'} -{\xt\over\xt'}\right ].}}
To this we must add the contributions from the singlets $P_i$, $\widetilde P_j$, $M_i$, $M'_j$.  Each of these fields couples only via a superpotential term in \eqnfourteen\ and, initially taking that singlet's
R-charge to be $2/3$, that superpotential term may be relevant or
irrelevant in the IR.  If it's relevant, then the singlet's R-charge is determined by the requirement that the
superpotential term have $R=2$ total in the IR.  If it's irrelevant, the singlet is a free field, with $R=2/3$.  
 If we assume that the last $p_P$ $P_j$'s (and ${\widetilde
P}_j$'s), the last $p_M$ $M_j$'s, and the last $p_{M'}$ $M'_j$'s
are interacting, then the quantity to maximize is 
\eqn\eqntwenty{\eqalign{
    \at^{(p)}/N_f'^2 &= \at^{(0)}/N_f'^2 + {2\nt\over 9}\sum _{j=1} ^{p_P} (2-3\alpha _j^P)^2(5-3\alpha _j^P)
    + {\nt^2\over 9}\sum _{j=1} ^{p_M} (2-3\alpha _j^M)^2(5-3\alpha _j^M)
    \cr &+
    {1\over 9}\sum _{j=1} ^{p_{M'}} (2-3\alpha _j^{M'})^2(5-3\alpha _j^{M'}) + 
   {4\nt\over 9}(k-2p_P)+
   {2 \nt^2\over 9}(k+1-2p_M) \cr &+ {2\over 9}(k+1-2p_{M'}),}}
where we define
 \eqn\eqnnineteen{\eqalign{
    \alpha _j^P &\equiv  \yt+{\xt\over\nt\xt'}(\yt-1) + {\xt^2\over\nt\xt'} - {\xt'\over\nt} + 1+(2j-1)\left [{1-\yt\over\xt'} +
    1-{\xt\over\xt'}\right],\cr
    \alpha _j^M &\equiv  2\left [{\xt\over\nt\xt'}(\yt-1) + {\xt^2\over\nt\xt'} - {\xt'\over\nt} + 1\right ]+2(j-1)\left [{1-\yt\over\xt'} +
    1-{\xt\over\xt'}\right],\cr
    \alpha _j^{M'} &\equiv 2\yt+2(j-1)\left [{1-\yt\over\xt'} +
    1-{\xt\over\xt'}\right].}}
The additional terms in \eqntwenty\ are the contributions from the singlets (see sect. 6 of \IWii\ for  a detailed discussion of an analogous example).  
The full solution $\yt (\xt, \xt ', \nt)$ is obtained by patching together the maximizing solutions of \eqntwenty\ with the appropriate values of $p_M$, $p_{M'}$, and $p_P$, depending on $(\xt, \xt', \nt)$, 
given by the largest integer $j$'s such that the $\alpha _j$ in \eqnnineteen\ satisfy $\alpha _j\leq 4/3$ (where the corresponding superpotential term becomes irrelevant).

For any given value of $k$, we can use the numerical a-maximization analysis to determine the range of $(\widetilde x, \widetilde x', \widetilde n)$, and thus the range of electric parameters $(x,x',n)$, in which the condition \wmagrel\ for $\Delta W=\Tr (Y\widetilde Y)^{k+1}$ to be relevant is satisfied.  Using \emratior, we'll express this
in terms of the electric parameters $(x,x',n)$.    To be concrete, let us consider the limit of large $k$.  The condition \welrel\ on the electric side gave the inequality \xminn, which shows that $x\approx x'$
must get large, linearly in $k$, in the large $k$ limit, while $n$ is restricted to the range \nrange.  
Then \emratior\ gives  $\xt \approx \xt'\approx k(1+n^{-1})-xn^{-1}$, and the condition \wmagrel\ will require $\xt$ to also be large.  In this limit of large $\xt \approx \xt '$, \eqnnineteen\ becomes
\eqn\eqntwentyeight{\eqalign{
    \at /N_f'^2 &\simeq 6\xt\left [1+{1\over\nt^2}\right ](\yt-1)^3 -
    20\xt(\yt-1) + {\xt \nt\over 36}\left (1+{1\over\nt}\right )\left [3(1-\yt)\left(1+{1\over\nt}\right
    )-4\right
    ]^3\cr
    &+{4\xt\over 9}\left[{\nt\yt\over 1-\yt}-1\right]+{\xt\nt\over 36}\left [{6\over \nt}(1-\yt)-4\right]^3+{2\nt\xt\over 9}\left[{\nt\over
    1-\yt}-2\right]\cr
    &+{\xt\over 36}\left[2-6\yt\right]^3+{2\xt(2\yt-1)\over 9(1-\yt)}+{2k\over
    9}(\nt+1)^2.}}
The first two terms are the large $\xt \approx \xt '$ limit of \eqnsixteen, and the rest 
are the remaining terms in \eqntwenty, with sums evaluated using \mesonsa\ (modifying 
the lower limit of the integral \mesonsa\ to be $2-3R=-2$, rather than $0$,
since $\alpha _j=4/3$ is the limit where the superpotential term becomes irrelevant).  
Maximizing \eqntwentyeight\ with respect to $\widetilde y$ gives $\widetilde y_{as}(\nt)$.

The condition \wmagrel\ for $\Tr (Y\widetilde Y)^{k+1}$ to be relevant can then be written for $k\gg 1$ as 
\eqn\magscwin{{(1-\widetilde y_{as})n\over k(n+1)-x}<{1 \over k}.} Rearranging and combining with 
\xminn, the electric and magnetic conditions \welrel\ and \wmagrel\ can be written together for $k\gg 1$ as 
\eqn\scwxx{1-y_{as}(n)<{x\over k}<1+n\widetilde y_{as}(n) \qquad\hbox{for $x\approx x'$ and $k\gg 1$.}}
For the duality \sunsund\ of \ILS\ to be useful, and the superconformal window be non-empty, the inequalities at the two ends of \scwxx\ had better be compatible with each other.  This is verified to indeed be the case, as seen in the plots in fig. 7, for the entire allowed range \nrange\ of $n$. The vertical axis of fig. 7 gives the allowed values of $x/k$, for a given value of $n$, and the superconformal
window is the region between the lower two curves on fig. 7.    There is also the stability bound 
conditions \stab, which in our $k\gg 1$ limit, with $x\approx x'$ scaling linearly in $k$, can both be written as $x/k<1+n$.  In the plot of fig. 7, the upper line is the stability bound, and the values of $x/k$ in the
superconformal window, between the lower two curves, is indeed always safely below the stability bound for the entire allowed range of $n$.  
All of these successes can be viewed as reassuring checks of the duality of \ILS.  
\bigskip
\centerline{\epsfxsize=0.40\hsize\epsfbox{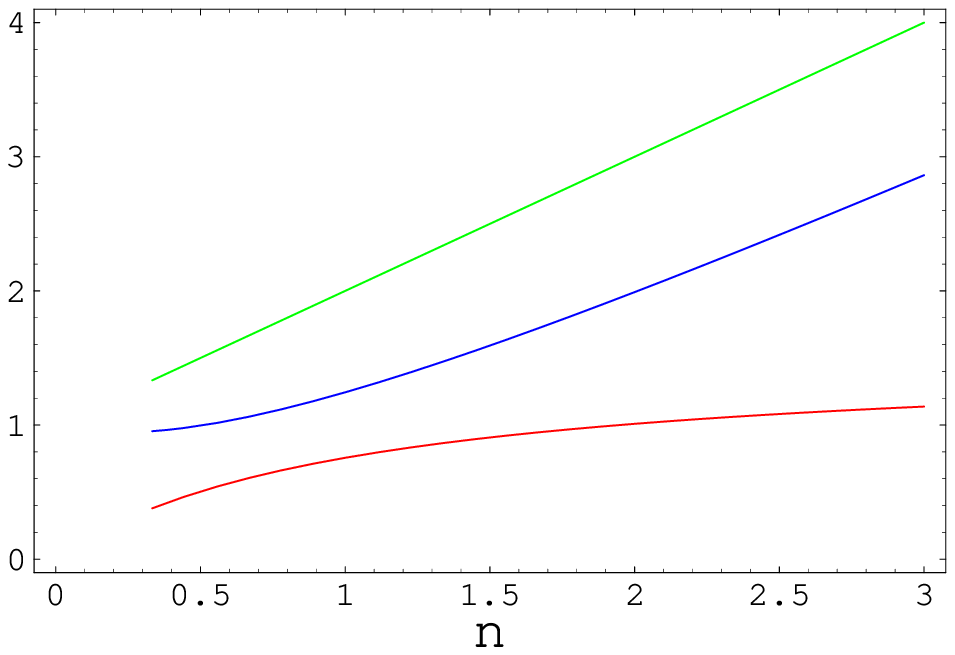}}
\centerline{\ninepoint\sl \baselineskip=2pt {\bf Figure 7:}
{\sl The $x/k$ conformal window: the upper line is the stability bound $1+n$,}}
\centerline{\ninepoint\sl the middle line is $1+n\widetilde y_{as}(n)$ and the lower line is $ 1-y_{as}(n)$. }
\bigskip
For the case $x=x'$,  $n=1$, the conformal window plotted in fig. 7 coincides with that obtained
in \KPS\ for SQCD with an adjoint, for the reason discussed above.

\newsec{Dualizing one gauge group}

%\KlebanovHB
\lref\KlebanovHB{
I.~R.~Klebanov and M.~J.~Strassler,
``Supergravity and a confining gauge theory: Duality cascades and
chiSB-resolution of naked singularities,''
JHEP {\bf 0008}, 052 (2000)
[arXiv:hep-th/0007191].
%%CITATION = HEP-TH 0007191;%%
}
%\CachazoSG
\lref\CachazoSG{
F.~Cachazo, B.~Fiol, K.~A.~Intriligator, S.~Katz and C.~Vafa,
`A geometric unification of dualities,''
Nucl.\ Phys.\ B {\bf 628}, 3 (2002)
[arXiv:hep-th/0110028].
%%CITATION = HEP-TH 0110028;%%
}

With product gauge groups, such as \sunsun, we can consider dualizing one of the gauge
groups, treating the other gauge group as a spectator.  The validity of doing this deserves
scrutiny, because duality is only exact at the IR fixed point.  Dualizing away from the extreme
IR can be potentially justified if the dualized group's dynamical scale is far above that of the 
other ``spectator" group, $\Lambda _d\gg \Lambda _s$ (and then holomorphic quantities can be analytically continued in $\Lambda_d/\Lambda _s$) or if one group gets strong while the other gets weak in the IR (as in string theory examples, see e.g. \refs{\KlebanovHB, \CachazoSG}).     

Let's consider the $SU(N_c)\times SU(N_c')$ theory \sunsun, and consider dualizing 
$SU(N_c)$, supposing that it's valid to treat $SU(N_c')$ as a weakly gauged flavor symmetry spectator. 
We'll suppose that the original electric theory satisfies \susuintir, so that both electric couplings
RG flow to non-zero values.  (If the second inequality \susuintir\
is violated, $SU(N_c')$ is IR free, and thus reasonably treated as a spectator.  But
if the first inequality in \susuintir\ is violated then $SU(N_c)$ is actually IR free, and the validity of
dualizing it with $SU(N_c')$ treated as a spectator is questionable.)  
The $SU(N_c)$ group has  $N_f+N_c'$ flavors and its Seiberg \NSd\ dual has $SU(\widetilde N_c)$ gauge group, with $\widetilde N_c\equiv N_f+N_c'-N_c$, with $N_f+N_c'$ flavors of dual quarks and $(N_f+N_c')^2$ singlet mesons.  The stability condition \susustabx\ ensures that $\widetilde N_c>0$.  Gauging $SU(N_c')_{mag}$, with the subscript as a reminder that its spectrum now differs from that of \sunsun, the dual is
\eqn\sundsun{\matrix{&&\hbox{gauge group:} &SU(\widetilde N_c)\times SU(N_c')_{mag}&\cr
&\hbox{matter:} &Y \oplus \widetilde Y&\bf{(\fund , \overline{\fund} )}\oplus \bf{(\overline {\fund}, \fund)},&\cr
&&q_f \oplus \widetilde q_{\tilde f}&\bf{(\fund , 1)} \oplus \bf{(\overline{\fund}, 1)}&(f, \widetilde f=1\dots N_f),\cr
&&Q'_{f'}\oplus \widetilde Q'_{\tilde f'}& \bf{(1, \fund)}\oplus \bf{(1, \overline{\fund})} &(f',\tilde f'=1\dots N_f'), \cr
&&F'_{n'}\sim X\widetilde Q\oplus c.c. & \bf{(1, \fund )}\oplus \bf{(1, \overline{\fund })} &(n',\tilde n'=1\dots N_f), \cr
&&M_{f, \tilde g}\sim Q\widetilde Q& \bf{(1, 1)}&(f,\tilde g=1\dots N_f), \cr
&&\Phi \sim X\widetilde X& \bf{(1, Adj)\oplus (1,1)},&\cr
}}
with the superpotential of \NSd\ yielding 
\eqn\wsutsup{
W=Mq\widetilde q+YF'\widetilde q+\widetilde Yq\widetilde F'+\Phi Y
\widetilde Y .}

The one loop beta function coefficients of the electric theory \sunsun\ were
\eqn\sunsunbeta{b_1=3N_c-N_c'-N_f, \quad\hbox{and}\quad b_1'=3N_c'-N_c-N_f'}
 (writing $b_1>0$ if asymptotically free), and those of the dual \sundsun\ are
\eqn\sundsunbeta{b_1{}^{mag}=2N_f+2N_c'-3N_c, \quad\hbox{and}\quad
b_1'{}^{mag}=N_c'+N_c  -2N_f-N_f'.} 
Note that $b_1'{}^{mag}$ differs from $b_1'$, because the $SU(N_c')_{mag}$
fields in \sundsun\ differ from those of the original $SU(N_c)\times SU(N_c')$ theory \sunsun; in fact, $b_1'-b_1'{}^{mag}=2(N_f+N_c'-N_c)=2\widetilde N_c>0$, so $SU(N_c')_{mag}$ is always less asymptotically free than the electric $SU(N_c')$ in the UV.  Ignoring the $SU(N_c')_{mag}$
dynamics, we'd conclude that the magnetic $SU(\widetilde N_c)$ is IR free if $N_f+N_c'<{3\over 2}N_c$; we'll discuss here how the $SU(N_c')_{mag}$ dynamics can dramatically affect when
the magnetic group is actually IR free.  

The important quantities for the IR dynamics are the exact beta functions for the theory \sundsun, which  are
\eqn\onednsvz{\beta _{g_{mag}}=-{3g_{mag}^3f \over 16\pi ^2}\Tr ~ SU(\widetilde N_c)^2 R, \quad
\beta_{g'_{mag}}=-{3g'{}^3_{mag}f'\over 16\pi ^2}\Tr ~SU(N_c')^2_{mag}R,}
where again $f$ and $f'$ are unimportant, positive, scheme dependent factors.  The beta functions \onednsvz\ can be written in the usual NSVZ form using \dimrc, which gives
\eqn\onednsvzz{\eqalign{3\Tr ~SU(\widetilde N_c)^2R&=b_1^{mag}+N_c'\gamma _Y+N_f\gamma _q, \cr
3\Tr ~SU(N_c')_{mag}^2R&=b_1'^{mag}+\widetilde N_c\gamma _Y+N_f'\gamma _{Q'}+N_f\gamma _{F'}+
N_c'\gamma _\Phi.}}

As with the electric theory, the dual \sundsun\ has three possible RG fixed points, 
\eqn\drgfpa{(\widetilde A)\ g_{mag*}\neq 0,\ g'_{mag*}=0, \quad\hbox{i.e. $SU(N_c')_{mag}$ free and $\Tr \, SU(\widetilde N_c)^2R|_{\widetilde A}=0$,}}
\eqn\drgfpb{(\widetilde B)\ g_{mag*}=0, \ g'_{mag*}\neq 0, \quad\hbox{i.e. $SU(\widetilde N_c)$ free 
and $\Tr \, SU( N_c')_{mag}^2R|_{\widetilde B}=0$,}}
\eqn\drgfpc{(\widetilde C)\ g_{mag*}\neq 0, g'_{mag*}\neq 0 \quad\hbox{so $\Tr \, SU(\widetilde N_c)^2R|_{\widetilde C}=\Tr \, SU( N_c')_{mag}^2R|_{\widetilde C}=0$,}}
which are depicted in fig. 8.  RG fixed point $(\widetilde A)$ is simply the Seiberg dual description of RG fixed point $(A)$ of the original electric theory (with $SU(N_c')$ part of the global flavor symmetry).
We expect that RG fixed point $(\widetilde C)$, when it exists,  is an equivalent, dual description of
the SCFT at RG fixed point $(C)$ of the original electric theory \sunsun.  The qualifier ``when it exists" is
because, as in the electric description, the RG flow may look like that of fig. 2 rather than that of fig. 1.
In the electric description, the condition for the RG fixed point (C)  to exist is \bzexisint.  We will determine
its analog in the magnetic theory \sundsun, for $(\widetilde C)$ to exist,  by analyzing the IR stability of the RG fixed points
$(\widetilde A)$ and $(\widetilde B)$ to small non-zero perturbations in the couplings that are set 
to zero in  \drgfpa\
and \drgfpb, in analogy with fig. 5.   
We will find that, for a particular range of flavors and colors, the theory
\sundsun\ with superpotential \wsutsup\ realizes the RG flow depicted in fig. 3: even if the one-loop beta
function might suggest that $SU(\widetilde N_c)$ is IR free, it can be driven to be interacting by the
interactions of the other gauge group and the superpotential.    
\bigskip
\centerline{\epsfxsize=0.80\hsize\epsfbox{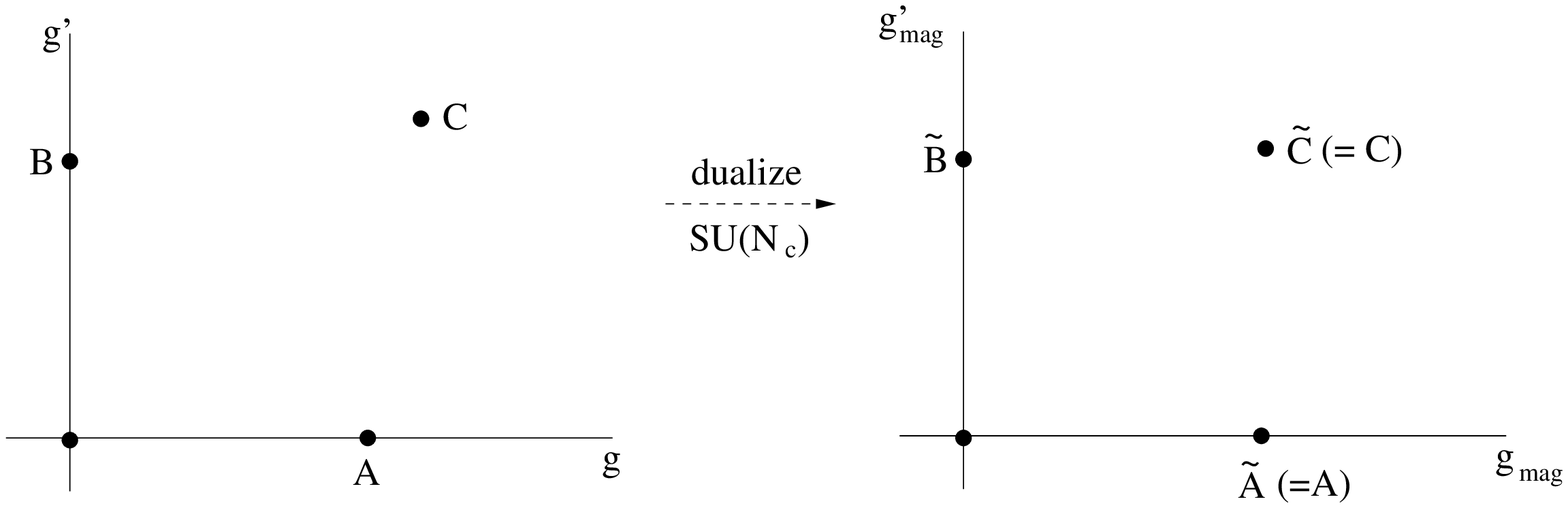}}
\centerline{\ninepoint\sl \baselineskip=2pt {\bf Figure 8:}
{\sl The process of dualizing one group.}}
\bigskip

Finally, we note that RG fixed point $(\widetilde B)$ is {\it not} the dual of RG fixed point (B): as duality exchanges strong and weak coupling, the RG fixed point $(\widetilde B)$, where the magnetic
$SU(\widetilde N_c)$ is free, corresponds to strongly coupled electric $SU(N_c)$.     In cases where
RG fixed point $(\widetilde B)$ is IR stable, our interpretation is that the electric side appears to
flow to interacting RG fixed point (C), but the magnetic dual reveals that the theory actually flows
instead to having a free magnetic $SU(\widetilde N_c)$, at the RG fixed point $(\widetilde B)$.

\subsec{The RG fixed point  $(\widetilde A)$ and its IR stability to $g'_{mag}$ perturbations.}

RG fixed point $(\widetilde A)$ is the Seiberg dual description of RG fixed point $(A)$ of
the original electric theory.  All of the superconformal R-charges at $(\widetilde A)$ are immediately
computable from the dual matter content \sundsun\ and superpotential, or from the Seiberg duality map \NSd\ and the superconformal R-charges
at RG fixed point $(A)$ in the electric description.  Either way, the result is:
$R(Y)=R(q)=N_c/(N_f+N_c')$, $R(M)=R(F')=R(\Phi)=2-2N_c/(N_f+N_c')$, and $R(Q')=2/3$. Using
\onednsvz, we see that $g'_{mag}$ is an IR relevant perturbation of $(\widetilde A)$ if $\Tr \, SU(N_c')_{mag}^2R|_{\widetilde A}$ is positive, or an IR irrelevant perturbation if it's negative.  This
't Hooft anomaly is easily directly computed, or we can use the fact that 't Hooft anomalies match
in Seiberg duality \NSd\ (since $SU(N_c')$ is a subgroup of the flavor group), so 
 $\Tr \, SU(N_c')_{mag}^2R|_{\widetilde A}=\Tr\, SU(N_c')^2R|_A$.
The RG fixed point $(\widetilde A)$ of the dual theory is thus IR stable if precisely the same inequality
\rgfpas\ found in the electric description holds.  So our first necessary condition for RG fixed point $(\widetilde C)$ to exist, at least within the domain of attraction of the UV fixed point at zero couplings, is that the  opposite inequality of \rgfpas\ should hold, 
 \eqn\betaprimexx{(3N_c'-N_f')(N_c'+N_f)-3N_c^2>0,}
to have $(\widetilde A)$ be IR repulsive.  
It is satisfying to see that the magnetic $(\widetilde A)$ RG fixed point is IR repulsive precisely when the
electric RG fixed point $(A)$ is.  It is hard to imagine how it could have been otherwise, given that the
RG fixed points $(A)$ and $(\widetilde A)$ are identified.

\subsec{The RG fixed point  $(\widetilde B)$ and its IR stability to $g_{mag}$ perturbations.}

This case is considerably more difficult than that of the previous subsection, as a-maximization
is needed to determine the superconformal R-charges at RG fixed point $(\widetilde B)$.  Notice that,
with the $SU(\widetilde N_c)$ gauge coupling set to zero at $(\widetilde B)$, the $SU(N_c')_{mag}$ 
spectrum in \sundsun\ is the same as that analyzed in \KPS: SQCD with an additional adjoint.  
But here the a-maximization analysis is further complicated by the presence of the 
superpotential in \sundsun, which couples some of the $SU(N_c')_{mag}$ fundamentals $Y$ to the adjoint $\Phi$, and also to fundamentals $F'$ and $SU(N_c')_{mag}$ singlets
$q$.  Rather than maximizing $a_{trial}$ as a function of one variable, $y$, depending on one parameter, $x$, as in \KPS, we'll have here to maximize $a_{trial}$ as  function of two variables, $R(Q')\equiv u$ and $R(\Phi)\equiv v$, depending on the three parameters $(x,x', n)$ of \ratios.  

Let's consider the constraints on the superconformal $U(1)_R$ at $(\widetilde B)$.  Having $\beta _{g'_{mag}}=0$ requires $\Tr\, SU(N_c')^2_{mag}R|_{\widetilde B}=0$ \drgfpb:
\eqn\suntsunpaf{N_c'+ \widetilde N_c(R(Y)-1)+N_f'(R(Q')-1)+N_f(R(F')-1)+N_c'(R(\Phi )-1)=0.}
The superpotential terms \wsutsup\ further impose 
\eqn\suntsunpwr{R(Y)+R(F')+R(q)=2, \quad \hbox{and}\quad
R(\Phi)+2R(Y)=2.}
Note that the first term in the superpotential \wsutsup\ is irrelevant for $g_{mag}=0$, since none of its fields are charged
under $SU(N_c')$, so $M$ is a free field, with $R(M)=2/3$.  The constraints \suntsunpaf\ and
\suntsunpwr\ are three constraints on five R-charges; they can be solved for 
\eqn\suntsunsf{R(Y)=1-\half v, \quad R(q)=\half (x+x')v+n(u-1), \quad R(F')=1+n(1-u)+\half (1-x-x')v,}
with $R(Q')\equiv u$, $R(\Phi)\equiv v$, and $(x,x',n)$ defined as in \ratios.  a-maximization w.r.t. $u$ and $v$ is needed
to determine  the values of $u(x,x',n)$ and $v(x,x',n)$.

Once we've determined the superconformal R-charges at $(\widetilde B)$, we can determine
whether or not $(\widetilde B)$ is stable to non-zero $g_{mag}$ perturbations.  We see from $\beta _{g_{mag}}$ in \onednsvz\ that $g_{mag}$ is a relevant perturbation of  $(\widetilde B)$ if $\Tr\, SU(\widetilde N_c)^2R|_{\widetilde B}>0$, i.e. if 
\eqn\sutsupr{3\widetilde N_c-N_f-N_c'+N_f\gamma _q+N_c'\gamma _Y>0,\quad\hbox{i.e. if}\quad 
- N_c+N_fR(q)+N_c'R(Y)>0.}
This condition, together with \betaprimexx, are the necessary conditions for RG fixed point
$(\widetilde C)$ to exist (at least within the domain of attraction of zero couplings).  If the 
inequality in \sutsupr\ is not satisfied, RG fixed point $(\widetilde B)$ is IR attractive, and then we expect RG flows from generic values of the couplings to end up there in the IR.  So if \sutsupr\ is not satisfied, the original electric theory \sunsun\ flows to having a free magnetic $SU(\widetilde N_c)$ in the IR.  

The condition \sutsupr, for $SU(\widetilde N_c)$ to not be free magnetic in the IR, is generally very different from the naive criterion, $N_f+N_c'>{3\over 2}N_c$, based on when $SU(\widetilde N_c)$
is asymptotically free in the UV.  The difference is that \sutsupr\ accounts for the $SU(N_c')_{mag}$
dynamics.  If the numbers of flavors and colors are chosen such that the $SU(N_c')_{mag}$ matter spectrum is just barely asymptotically free (i.e. $b_1'{}^{mag}$ in \sundsunbeta\ is small and positive),
then the RG fixed point coupling $g'_{mag}$ at $(\widetilde B)$ is small.  In this case, the
the $SU(N_c')_{mag}$ dynamics doesn't much affect the running of the $SU(\widetilde N_c)$ coupling $g_{mag}$.  In particular, when $(\widetilde B)$ is at weak coupling, the a-maximization results
properly give $R(q)\approx 2/3$ and $R(Y)\approx 2/3$, since these fields are approximately free.
We then find that \sutsupr\ gives approximately the standard condition from Seiberg duality \NSd\
for the magnetic dual $SU(\widetilde N_c)$ to be interacting rather than IR free, $N_f+N_c'>{3\over 2}N_c$, which is the condition that $g_{mag}$ be an IR relevant perturbation of  free theory at $g_{mag}=g'_{mag}=0$.  

On the other hand, when the number of flavors and colors are such that $SU(N_c')_{mag}$ is
very much asymptotically free, i.e. $b_1'{}^{mag}$ in \sundsunbeta\ is positive and large, the 
RG fixed point $(\widetilde B)$ is at strong $SU(N_c')_{mag}$ coupling.  In this case, the
$SU(N_c')_{mag}$ dynamics can radically affect whether or not the $SU(\widetilde N_c)$
coupling $g_{mag}$ is relevant.  Indeed,  depending 
on the values of $(x,x',n)$,  this theory can realize the flow of fig. 3: even if $N_f+N_c'\leq {3\over 2}N_c$,
so $SU(\widetilde N_c)$ is IR free around $g_{mag}=g'_{mag}=0$,
the condition \sutsupr\ for $SU(\widetilde N_c)$ to be an IR interacting deformation of $(\widetilde B)$  can nevertheless be satisfied.  In short, the $SU(N_c')_{mag}$ has driven an otherwise IR free
$SU(\widetilde N_c)$ theory to instead be IR interacting.    We'll focus on this phenomenon of fig. 3
in the rest of this subsection.  

To have \sutsupr\ be satisfied when $b_1^{mag}<0$ requires that $q$ and/or
$Y$ have {\it positive} anomalous dimension, i.e. R-charge greater than 2/3, at the RG fixed point (B).  
Positive anomalous dimensions are possible provided that there is a $W_{tree}$ superpotential,
as in the case here \wsutsup: gauge interactions make negative contributions to the anomalous
dimensions, and superpotential interactions make positive contributions.  As we'll now discuss, there is indeed a range of
 parameter space of flavors and colors for the  magnetic theory \sundsun\ where the anomalous dimensions are sufficiently positive so as to have  \sutsupr\ satisfied,
 despite having $b_1^{mag}<0$, realizing the effect of fig. 3. (See also the example \CMT.)

Though the a-maximization analysis at RG fixed point $(\widetilde B)$ is standard, it's computationally intensive.  Using the spectrum 
\sundsun\ and \suntsunsf, we compute the combination of 't Hooft anomalies $a^{(0)}=3\Tr R^3-\Tr R$, 
as a function of the two variables, $(u,v)$, and the parameters $(x,x',n)$.  Depending on $(x,x',n)$,
we have to also add the additional contributions \acorr\ for any gauge invariant operators with
$R\leq 2/3$.  (The operators hitting the unitarity bound are found to be $Q'\Phi ^{j-1}\widetilde Q'$, $Q'\Phi ^{j-1}\widetilde F'$, and $F'\Phi ^{j-1}\widetilde F'$ for values of $j=1\dots $ increasing with $x$ and $x'$, 
as the RG fixed point $(\widetilde B)$ becomes more and more strongly coupled.) 
Again, we implemented this a-maximization analysis numerically, using Mathematica.
While the numerics are similar in spirit to our previous cases, the fact that here we're maximizing a 
function of two, rather than one, variables, as a function of the three parameters, greatly prolongs the
required computational timescale.  

Let's focus on an interesting range of parameter space, where we take the parameters $x$ and
$x'$ to be large.  This is an interesting region of parameter space because then RG fixed point
$(\widetilde B)$ is at very strong $SU(N_c')_{mag}$ gauge coupling (as seen from the fact that the
one-loop beta function is very large).    For large $x$ and $x'$, there are terms quartic in $x$ in 
$a^{(0)}=3\Tr R^3-\Tr R$, coming from the $O(x)$ terms in $R(q)$ and $R(F')$ in \suntsunsf.  Note,
however, that the $O(x)$ terms in \suntsunsf\ all appear multiplied by $v$, so the leading terms for
large $x$ transform homogeneously, with degree one, under $x\rightarrow \lambda x$ and $v\rightarrow \lambda ^{-1}v$, e.g. the quartic term
in $a^{(0)}$ is $\sim x^4v^3$.  When we include the contributions from the meson hitting the unitarity bound, we find that they also have a leading large $x$ term which is degree one under this scaling. 
 Because of this homogeneity, when we maximize w.r.t. $v$ for large $x$, we find $v\sim 1/x$,
and the value of $a$ at its maximum is linear in $x$, since it's degree one in the above rescaling.  (The
fact that the central charge of the SCFT grows linearly in $x$ is a check of the a-theorem conjecture,
since there's a Higgsing RG flow analogous to that of \KPS\ which would violate the a-theorem with 
any higher degree.)    To study the limit of large $x$, we can thus scale $\lambda \rightarrow \infty$, keeping only the terms of degree one in $\lambda$.

In this scaling limit, $R(\Phi)\equiv v\rightarrow \lambda ^{-1}v\rightarrow 0$ for $\lambda \rightarrow \infty$.  Then
\suntsunsf\  gives $R(Y)\rightarrow 1$, while $R(Q')\equiv u$, $R(q)$, and $R(F')$ asymptote to some
finite values that are determined by a-maximization.  The a-maximization answer for these quantities,
in our limit of large $x$ and $x'$, can also be obtained by borrowing the a-maximization results of
\KPS.  The idea is that the $SU(N_c')_{mag}$ gauge coupling at $(\widetilde B)$ is very strong for large
$x$ and $x'$, and the $SU(N_c')_{mag}$ matter content that it couples to coincides with that of \KPS,
and adjoint and fundamentals.   The theory at $(\widetilde B)$ differs from that of \KPS\ only because
of the superpotential interactions \wsutsup.  If not for the superpotential interactions, the a-maximization result of
\KPS\ would tell us that $R(Y)$, $R(Q')$ and $R(F')$ all asymptote to $(\sqrt{3}-1)/3\approx 0.244$ for
$x\rightarrow \infty$.    The superpotential non-negligibly affects $R(Y)$: the 
$\Phi Y\widetilde Y$ term requires that $R(Y)\rightarrow 1$, since $R(\Phi)\rightarrow 0$ for large $x$.  But the superpotential negligibly affects $R(Q')$ and $R(F')$ in the strong $SU(N_c')_{mag}$ coupling, large $x$ limit.  For example, though there is a term $YF'\widetilde q$ in the superpotential, its effect is to determine the R-charge of the 
otherwise free field $q$, leaving $R(Y)\rightarrow 1$ and $R(F')\rightarrow (\sqrt{3}-1)/3\approx 0.244$
unaffected.   So in this large $x$ limit we obtain (and the detailed a-maximization analysis bears this out):
\eqn\ryqlim{R(Y)\rightarrow 1, \quad \hbox{and}\quad R(q)\rightarrow 1-\left({\sqrt{3}-1\over 3}\right)\approx 0.756, \quad\hbox{for large $x$ and $x'$}.}  

The condition \sutsupr\ for $g_{mag}$ to be relevant at $\widetilde B$ is then (recalling \ratios)
\eqn\sutsuprr{N_c'-N_c+N_f(0.756)>0,\quad\hbox{i.e.}\quad x'-x  +0.756>0, \quad\hbox{for large $x$ and $x'$}.}
This is very different from the condition that $g_{mag}$ be asymptotically free for $g'_{mag}=0$, 
$b_1^{mag}>0$, i.e. $1+x'-{3\over 2}x>0$, and \sutsuprr\ can be satisfied even when $b_1^{mag}<0$, i.e.
we can have 
\eqn\figiiiex{1+x'-{3\over 2}x<0 \quad\hbox{but nevertheless}  \quad x'-x+0.756>0;}
for example, we can take $x\approx x'\rightarrow \infty$.  For values of $(x,x',n)$ such that 
the inequalities in \figiiiex\ both hold, the RG flow is as in fig. 3: if $SU(N_c')_{mag}$'s coupling were set to exactly
zero, then $SU(\widetilde N_c)$ would be IR free, but any non-zero $SU(N_c')_{mag}$ coupling
would eventually drive $SU(\widetilde N_c)$ to be instead interacting in the IR.  

\subsec{The RG fixed point $(\widetilde C)$}

RG fixed point $(\widetilde C)$ exists if  $(\widetilde A)$ and $(\widetilde B)$ are both IR unstable to
perturbations in the other coupling; we found this to be the case if 
 \betaprimexx\ holds, and if, say for  large $x$ and $x'$,  \sutsuprr\ holds, respectively.   When $(\widetilde C)$ exists, we expect that it's an equivalent, dual description of the RG fixed point $(C)$ of the original electric theory.  We'll here check that the superconformal
R-charges are compatible with this identification.

At $(\widetilde C)$, the six independent superconformal R-charges, on the six lines of \sundsun, are
subject to five constraints: for vanishing $\beta _{g_{mag}}$ and $\beta _{g'_{mag}}$, we require 
\drgfpc,  $\Tr \, SU(\widetilde N_c)^2R|_{\widetilde C}=\Tr \, SU( N_c')_{mag}^2R|_{\widetilde C}=0$, 
along with three more constraints from requiring that the superpotential terms
\wsutsup\ all have total $R(W)=2$.  (All terms in \wsutsup\ are relevant deformations of the $W=0$
theory when $g_{mag*}$ and $g'_{mag*}$ are both non-zero.)   There is thus a one-variable family 
of R-charges, as for the electric RG fixed point $(C)$.  These constraints are compatible with the 
duality map identification of the fields $F'$, $M$, and $\Phi$ in \sundsun:  the R-charges of the dual
theory \sundsun\ can be related to those of the original electric theory \sunsun, with $R(Q)\equiv y$ as before, by
\eqn\dualmr{R(M)=2y, \ R(F')=R(X)+y,\ R(\Phi )=2R(X),\ R(q)=1-y,\ R(Y)=1-R(X),}
with $R(X)$ and $R(Q')$ given by \rxyp\ in  terms of the variable $y$ and parameters $(x,x',n)$.  

We compute the same function $a^{(0)}_{trial}=3\Tr R^3-\Tr R$ to maximize w.r.t. $y$ as in the electric
theory \aosusu, as expected from the 't Hooft anomaly matching for the global flavor symmetries in
Seiberg duality \NSd.  Compatible with our claim that the electric RG fixed point $(C)$ is 
equivalent to the dual one $(\widetilde C)$, there is a one-to-one mapping of the operators that have hit the unitarity bound.  Corresponding to the operators \susumesons\ we have 
\eqn\mesondmap{M_{j=1}\leftrightarrow M,\ M_{j>1}\leftrightarrow \widetilde F
\Phi ^{j-2}F, \ M_j'\leftrightarrow Q'\Phi ^{j-1}\widetilde Q', P_j\leftrightarrow  F'\Phi ^{j-1}\widetilde Q',\ 
\widetilde P_j\leftrightarrow \widetilde F'\Phi ^{j-1}Q'.}
So, even including the contributions of the operators hitting their unitarity bound, we find the
same $a_{trial}$ function of $y$ and $(x,x',n)$ to maximize w.r.t. $y$, and hence the same superconformal R-charges are given by \dualmr\ with $y(x,x',n)$ the same superconformal
R-charge as obtained by analyzing the electric theory \sunsun.

\newsec{Conclusions and Comments}

A general potential pitfall in applying a-maximization is that one must really have the full
symmetry group under control, including all accidental symmetries, to obtain correct results.
Overlooking some symmetries will lead to a value of the central charge $a_{SCFT}$
that is too low.  Seiberg duality \NSd\ shows that there can be highly non-obvious accidental
symmetries, such as those acting on the free magnetic $SU(N_f-N_c)$ quarks and gluons
when $N_f<{3\over 2}N_c$.    More generally, without knowing the dual, we do not presently have a way to  look for such accidental symmetries, which do not act on any of the ``obvious" gauge invariant  operators of the theory.  

Ignoring the interplay of the two gauge couplings, the superconformal window of \NSd\ for
each gauge group in \sunsun\ separately is
\eqn\sususcwindn{{3\over 2}N_c<N_f+N_c'<3N_c, \qquad {3\over 2}N_c'<N_f'+N_c<3N'_c.}
These are the conditions for points (A) and (B) to be interacting SCFTs, respectively.  The
upper limits are needed for the electric coupling to not be driven to zero in the IR, and the lower
limits are for the couplings of the dual \NSd\ to not be driven to zero in the IR.  
 
Accounting for the $g$ and $g'$ interplay, the conditions for point (C) to exist as a fully 
interacting SCFT differ from \sususcwindn.  The upper limits of \sususcwindn\ should be
replaced with the conditions \bzexisint, for neither electric gauge coupling to be driven to 
zero in the IR.  Similarly, the duality of sect. 4 (assuming its validity) shows how the lower limits of \sususcwindn\ are modified, in order for neither magnetic gauge coupling to be driven to zero
in the IR.  

For example, taking $x\equiv N_c/N_f$ and $x'\equiv N_c'/N_f$ large, we found in sect. 4 that $(\widetilde B)$, with $g_{mag}\rightarrow 0$,  is IR attractive if 
\eqn\rgfpbds{N_c'-N_c+(0.756)N_f<0, \quad\hbox{i.e. if}\quad x'-x+0.756<0.}
In this case,  rather than flowing to the fully interacting RG fixed point (C), the theory flows to
the free magnetic point  $(\widetilde B)$ in the IR, where the original electric $SU(N_c)$ is very strongly coupled, but its $SU(N_f+N_c'-N_c)$ magnetic dual is IR free.  There is then  
a large, non-obvious, accidental symmetry of the  original electric theory when \rgfpbds\ holds.  
Likewise, dualizing the $SU(N_c')$ factor of \sunsun, we find for large $x$ and $x'$ that the 
apparent RG fixed
point $(C)$ of the electric theory instead flows to having a free magnetic $SU(N_f'+N_c-N_c')$ group when 
\eqn\rgfpbbds{N_c-N_c'+(0.756)N_f'<0, \quad\hbox{i.e. if}\quad x-x'+(0.756)n<0.}

So, for RG fixed point (C) to be fully interacting, rather than partially free magnetic, the lower limits in \sususcwindn\ are replaced, for large $x$ and $x'$, with the conditions 
\eqn\rgfpnbs{-(0.756)n<x-x'<0.756.}
The range \rgfpnbs\ is a subset of the stability range \susustabx.  Outside of the range \rgfpnbs, there are non-obvious accidental symmetries.  Within the range \rgfpnbs, we have no
evidence for non-obvious accidental symmetries.  If there had been any such non-obvious 
accidental symmetries, our a-maximization analysis of sects. 2 and 3 would have to be 
appropriately modified.  In particular, in our parameter slice of special interest in sects. 2 and 3, $x=x'$, i.e. $N_c=N_c'$, the magnetic
duals remain fully interacting.  

As we noted, for $x=x'$ and $n=1$, the a-maximization analysis of our product group example \sunsun\ coincides 
with that in \KPS\ for $SU(N_c)$ with an adjoint and $N_f$ fundamentals.  In the analysis
of \KPS\ of that latter theory, it was assumed that the only accidental symmetries are the obvious
ones, associated with gauge invariant operators hitting the unitarity bound.  But, as in the 
example of \CMT, there's a possibility of a non-obvious accidental symmetry, associated with
a free-magnetic gauge group in a deconfining dual.  The idea of the deconfining dual \Berkooz\
is that the dual \sunsund\ of our theory \sunsun\ would look quite a lot like 
$SU(N_c')$ SQCD with an adjoint if we chose the flavors and colors such that $\widetilde N_c=1$ in \sundsun.  And a slight modification of the theory \sunsun, with added 
fields and superpotential terms (designed to eliminate the analog of \wsutsup\ in the dual), will lead to precisely SQCD with an adjoint and fundamentals, with no superpotental; see table 8 of \Luty\ 
for the needed field content.     It would be interesting  to carry out the a-maximization analysis of that theory, and its duals, to determine whether or not any of the gauge groups of the deconfining duals
can become IR free.  

\centerline{\bf Acknowledgments}

We would like to thank M.T. Fleming for discussions and participation in the early stage of
this work. We also thank M. Schmaltz,  W. Skiba, and R. Sundrum  for discussion.  B.W. also acknowledges the hospitality of the KITP and thanks
the organizers of the QCD and Strings Workshop. 
This work was supported by DOE-FG03-97ER40546.  The work of BW was also 
supported in part by National Science Foundation Grant PHY-00-96515 and by
DOE contract DE-FC02-94ER40818.

\appendix{A}{On the superconformal window of the other duals of \ILS}

\subsec{Reviewing $SU(N_c)$ SQCD, with $N_f$ fundamental flavors, and an adjoint $X$}
Let us briefly review the a-maximization analysis of Kutasov, Parnachev, and Sahakyan (KPS) \KPS\ for this theory, with $W_{tree}=0$.   
The anomaly free superconformal R-charges of the fields are $R(Q)=R(\widetilde Q)\equiv y$
for the fundamentals and $R(X)=(1-y)/x$ for the adjoint, where $x\equiv N_c/N_f$. 
a-maximization determines $y(x)$ via maximizing 
\eqn\akpsi{\eqalign{a^{(p)}_{KPS}(y,x)/N_f^2 &=2x^2+x^2\left(3 \left({1-y\over x}-1\right)^3- \left({1-y\over x}-1\right)\right)+2x\left(3(y-1)^3-(y-1)\right)
\cr &+\sum _{j=0}^p\left(2y+j{1-y\over x}-{2\over 3}\right)^2\left(5-3[2y+j{1-y\over x}]\right),}}
w.r.t. $y$; this has solution $y^{(p)}(x)$.  The sums account for the generalized mesons $QX^j\widetilde Q$ hitting their unitarity bound, with $p$ the greatest integer such that $R(QX^j \widetilde Q)$ would naively violated the unitarity bound.  The solution $y_{KPS}(x)$ is obtained by patching together the functions $y^{(p)}(x)$, with the appropriate value of $p$ depending on $x$. 
The function $y_{KPS}(x)$ is monotonically decreasing, with asymptotic value $y(x\rightarrow \infty)\rightarrow y_{as}=(\sqrt{3}-1)/3$.  $R(X) = (1-y)/x$ is also monotonically decreasing in $x$ and, for $x\rightarrow \infty$, $R(X)\rightarrow (1-y_{as})/x$.  

The superpotential $W_{A_k}=\Tr X^{k+1}$ is a relevant deformation of the $W=0$ SCFT if $R(X^{k+1})<2$ (since $\Delta (W)={3\over 2}R(W)$), i.e. if
$R(X)=(1-y)/x<2/(k+1)$.   Since $R(X)$ monotonically decreases with $x$, $W_{A_k}$ can always be made relevant, by taking
$x$ sufficiently large, $x>x^{min}_{A_k}$, where $x^{min}_{A_k}$ is determined by the condition
that $(1-y(x^{min}_{A_k}))/x^{min}_{A_k}=2/(k+1)$.  Using the numerical solution for $y(x)$, the
numerical values of $x^{min}_{A_k}$ can be obtained for arbitrary $k$.  For large $k$, $x^{min}_{A_k}$ is large, and then $R(X)\approx (1-y_{as})/x^{min}_{A_k}=2/(k+1)$ gives 
$x^{min}_{A_k}
\rightarrow \left({4-\sqrt{3}\over 6}\right)k\approx 0.3780k$ \KPS.

The $A_k$ theory has dual description \kdual\ in terms of a magnetic
$SU(\ncd)$ gauge theory, with $\ncd =
k\nf-\nc$.   It has an adjoint field $Y$, $\nf$ fundamental flavors $q$,
$\tilde q$, and $N_f^2$ gauge singlet fields $M_j$, for $j=1\dots k$. The
superpotential is
\eqn\WDsuadj{
W_{\widetilde{A_k}}=\Tr Y^{k+1}+\sum _{j=1}^{k}M_{k-j}qY^{j-1}\tilde q \ .}
The analysis of the dual theory is similar to that of the electric theory, with $x\rightarrow
\widetilde x =\ncd/\nf=k-x$, though the specifics are not
identical, because of the effect of the additional gauge singlets $M_j$ and superpotential
terms in \WDsuadj.    We refer the reader to \KPS, for the details of the a-maximizing $\widetilde y(\widetilde x)$ in the magnetic theory.  The qualitative result is that $\widetilde y(\widetilde x)$
drops to zero a little faster on the magnetic side than the electric $y(x)$, so
the $\Tr Y^{k+1}$ term in \WDsuadj\
is relevant for $\widetilde x>\widetilde x_{A_k}^{min}$, with $\widetilde{x}_{A_k}^{min}<
x^{min}_{A_k}$.  In particular, for $k\gg 1$, $\widetilde{x}_{A_k}^{min}\approx 0.3578k$
\KPS .

The superconformal window, where both dual descriptions of the $A_k(N_c, N_f)$ SCFTs are useful, is $x_{A_k}^{min}<x<
k-\widetilde{x}_{A_k}^{min}$; for $k\gg 1$, it's  $0.3780k<x<0.6422k$. 
Within this range the electric and magnetic theories have
the same central charge $a$, as guaranteed by 't Hooft anomaly matching.  Outside of this
range, there are accidental symmetries that are manifest in one of the dual descriptions
but not in the other so, without accounting for these accidental symmetries, the central
charge as computed by a-maximization for the electric and magnetic theories can appear to
differ \KPS.  The larger of $a_{elec}$ or $a_{mag}$ is the correct one -- it's larger because
of maximizing $a_{trial}$ over R-symmetries that can mix with the additional, accidental
flavor symmetries.

\subsec{Some immediate generalizations, with other groups and matter content}

Many analogs of the duality of \kdual\ were soon given in \refs{\kispso, \rlmsspso, \ILS}, all
for theories with $W_{A_k}$ type LG superpotential.  Without the LG superpotential, those theories
are expected to flow to other SCFTs, which can now be analyzed via a-maximization.
Doing so determines when the $W_{A_k}$ superpotential is relevant.  Doing a similar a-maximization analysis
of the duals of  \refs{\kispso, \rlmsspso, \ILS}\ determines when the dual $A_k$ LG superpotentials are relevant.  Combining the two bounds, as in the analysis of \KPS, reviewed in the last section, determines the superconformal
window for where the duals of \refs{\kispso, \rlmsspso, \ILS}\ are useful.  In particular, we can verify that
the superconformal window is non-empty for all $k$.  

As in the analysis of \KPS, it is most convenient to consider the limit $N_c\gg 1$, $N_f\gg 1$, holding
$x\equiv N_c/N_f$ fixed.   However, as we'll now explain, the a-maximization analysis of all of the examples of \refs{\kispso, \rlmsspso, \ILS}\ involving a single gauge group becomes simply identical to that of \KPS\ in this limit, where we drop terms $O(1/N_c)$ or $O(1/N_f)$.

The examples of \refs{\kispso, \rlmsspso, \ILS}\ involving a single gauge group are:
\eqn\allsame{\matrix{\hbox{group}&X&Q&\hbox{\# mesons $QX^jQ$}&a/a_{KPS}\cr
\ SO(N_c)\ &\ \sym\  & \ N_f\cdot \fund \ &\ \half N_f(N_f+1)&\half \ \cr
Sp(N_c) & \asym & 2N_f \cdot \fund & N_f(2N_f-1) & 2\cr
\ SO(N_c)\ &\ \asym\  & \ N_f\cdot \fund \ &\ \half N_f(N_f+(-1)^j)&\half \ \cr
Sp(N_c) & \sym & 2N_f \cdot \fund & N_f(2N_f-(-1)^j) & 2  \cr
SU(N_c) & \asym \oplus \overline{\asym}& N_f\cdot (\fund \oplus \overline{\fund})& \hbox{$N_f^2$ or $N_f(N_f-1)$}
& 1\cr
SU(N_c) & \sym \oplus \overline{\sym}& N_f\cdot (\fund \oplus  \overline{\fund})& \hbox{$N_f^2$ or $N_f(N_f+1)$}
& 1\cr
SU(N_c) & \asym \oplus \overline{\sym}& 8\cdot \fund\oplus N_f\cdot (\fund \oplus \overline{\fund})& \hbox{$\sim N_f^2$}
& 1
\cr}}
Our notation for $Sp(N_c)$ is that $SU(2)\cong Sp(1)$.

Let us compare the theory on the first line of \allsame\ with the $SU(N_c)$ with adjoint
theory analyzed in \KPS.  The anomaly free R-symmetry is constrained to
satisfy $2(N_c-2)+2N_f(R(Q)-1)+2(N_c+2)(R(X)-1)=0$.  But in the $N_c\gg 1$ and
$N_f\gg 1$ limit, holding fixed $x\equiv N_c/N_f$, this gives an identical relation,
$R(X)=(1-y)/x$, with $x\equiv N_c/N_f$, as in the case reviewed in the previous subsection.
Computing the analog of \akpsi\ for the theory on the first line of \allsame, we find that
every term is now simply half of that in \akpsi, coming from the fact that the only difference
(to leading order $O(1/N_c)$ and $O(1/N_f)$) is that there are half as many of each of the
different fields.  For example, the $2x^2$ term in \akpsi\ is the contribution of the $|SU(N_c)|
\approx N_c^2$ gauge fields, which here becomes a similar contribution from the
$|SO(N_c)|=\half N_c(N_c-1)\approx \half N_c^2$ gauge fields.  Similarly, there are here
half as many $X$ fields ($\half N_c(N_c+1)\approx \half N_c^2$ here, vs. $N_c^2-1\approx N_c^2$ there),
half as many $Q$ fields ($N_cN_f$ here, vs $2N_cN_f$ there) and half as many of the meson fields
($\half N_f(N_f+1)\approx \half N_f^2$ here, vs $N_f^2$ there).  So the a-function to maximize for
the theory in the first line of \allsame\ is simply half $a_{KPS}$ \akpsi.  Maximizing this obviously leads to the same solution for the superconformal R-charges as obtained in \KPS, $y(x)=y_{KPS}(x)$.

Likewise, all of the other theories in \allsame\ similarly lead to the same results in the $N_c\gg 1$,
$N_f\gg 1$ limit, for arbitrary $x\equiv N_c/N_f$.  In this limit, the anomaly free condition
on the superconformal R-symmetry gives $R(X)=(1-y)/x$, with $R(Q)\equiv y$, for all of them.
For all of these theories, the analog of every term in \akpsi\ becomes approximately simply the same
as in \akpsi, up to an overall factor, which is given in the last column of \allsame.  For
example, for the theory in the last line of \allsame, the generalized mesons $QX^jQ$ which can hit
the unitarity bound are given for even $j$ by
$\widetilde Q(X\widetilde X)^r Q$ (which are $N_f(N_f+8)\approx N_f^2$ in number) and for odd $j$ by
$Q(\widetilde XX)^2 \widetilde X Q$ (which are $\half (N_f+8)(N_f+9)\approx \half N_f^2$ in number) and $\widetilde QX(\widetilde XX)^r \widetilde Q$ (which are $\half N_f(N_f-1)\approx \half N_f^2$ in
number) so, whether $j$ is even or odd, there are approximately the same number $N_f^2$ of mesons,
leading to the same contributions as in the last line of \akpsi.

There are several interrelations among the theories \allsame\ associated with giving $X$
an expectation value (which we're free to do, since we're now discussing the theories with
$W_{tree}=0$), and it can be verified that all of these Higgsing RG flows satisfy $a_{IR}<a_{UV}$.
These checks make use of the $a/a_{KPS}$ factors in the last column of \allsame.  For example, consider the $SO(N_c)$ theory with adjoint  $X$, on the third line of \allsame.   Giving $X$ an expectation
value, there is a RG flow connecting this $SO(N_c)$ theory in the UV to an IR theory with gauge
group $U(\half N_c)$, adjoint matter $X_{low}$,
and $N_f$ fundamental flavors $(\fund \oplus \overline{\fund})$.  Using the last column of \allsame,
the UV theory has $a_{UV}\approx \half N_f^2 \widehat a_{KPS}(x)$.  The IR theory has
$a_{IR}\approx N_f^2 a_{KPS}(\half x)$, because the IR theory has $N_c/2$ colors.  The a-theorem
conjecture thus requires $\half \widehat a_{KPS}(x)>\widehat a_{KPS}(\half x)$, which can be verified
to be satisfied.

Since all of the theories in \allsame\ have the same R-charge $R(X)$, given by $R(X)=(1-y_{KPS}(x))/x$,
the minimal values $x^{min}_{A_k}$ for the $W_{A_k}=\Tr X^{k+1}$ superpotential to be relevant
is the same, for all of these theories\foot{The theories in the third through sixth line of \allsame\ must have $k=odd$, and that in the last line of \allsame\ must have $k+1 =0 \ (mod)\ 4$.},
as was obtained in \KPS\ for the $SU(N_c)$ with adjoint theory.  E.g. for $k\gg 1$, all have $x^{min}_{A_k}\rightarrow \left({4-\sqrt{3}\over 6}\right)k$.  

We can similarly analyze the magnetic duals of the above theories \refs{\kispso, \rlmsspso, \ILS}.
For example, the theory in the last line of \allsame, upon deforming by superpotential $W_{A_k}=
\Tr (X\widetilde X)^{\half(k+1)}$ (with $k+1=0$ mod 4 here), was argued to be dual to a similar theory, with gauge group $SU(\widetilde N_c)$, with $\widetilde N_c\equiv k(N_f+4)-N_c$, along with some additional gauge singlets and superpotential terms.  We can use a-maximization to analyze this dual $SU(\widetilde N_c)$ theory for $W_{tree}=0$, and thereby determine when the various
terms in the superpotential appearing in the duality of \ILS\ are relevant.  In particular, we can determine
$\widetilde x^{min}_{A_{k}}$, the lower bound on $\widetilde x\equiv \widetilde N_c/N_f$ in order for
the superpotential $\widetilde W_{A_{k}}$ to be relevant.  As on the electric side, in the limit of
large $N_c$ and $N_f$, the a-maximization analysis becomes identical to that of \KPS\ for the magnetic dual of the adjoint theory with  superpotential $W_{A_{k}}$: the above $\widetilde N_c$ becomes
$\widetilde N_c\approx kN_f-N_c$, as in the adjoint theory, and every term in the a-maximization
analysis here maps to a corresponding term there.  In this limit, the values here of the $\widetilde x^{min}_{A_{k}}$ are the same as those obtained in \KPS\ for the adjoint theory.

Thus, at least in the $N_c\gg1$ and $N_f\gg 1$ limit, all of the above theories have exactly
the same superconformal window as obtained in \KPS\ for the $SU(N_c)$ theory with adjoint.

We also note that all of the other theories in \ILS, with product  gauge groups, also have the same superconformal window range of $x$, as long as we take all the groups to have the same (large) rank, and take all to have the same (large) number of fundamental flavors.   This generalizes our observation of Sect. 2,  that the 
$SU(N_c) \times SU(N_c')$ theory gives the superconformal window obtained in \KPS\ for 
the slice of  parameter space $N_c = N_c'$ and $N_f = N_f'$.  

For example, consider the $SU(M)
\times SO(N)\times SO(N')$ duality discussed in sect. 11 of \ILS.  In the parameter slice, 
$M=N=N'\equiv N_c$, and $m=n=n'\equiv N_f$, taking $N_c\gg 1$ and $N_f\gg 1$ large, holding
fixed the ratio $x\equiv N_c/N_f$, we find that every term in the quantity $a_{trial}(y,x)=3\Tr R^3-\Tr R$ equals twice a corresponding term in the corresponding function of \KPS,  $a_{trial}(y,x)=2a_{KPS}(y,x)$ (even including the contributions of the gauge invariant operators that hit the unitarity bound).   This is because there is a correspondence in this limit between every field, with the same R-charges
and twice as many copies for the $SU\times SO\times SO$ theory of \ILS\ as compared with that
of \KPS.    Since  $a_{trial}=2a_{KPS}$, it is  maximized by the same function, $y=y_{KPS}(x)$.  The anomalous dimensions are thus the same of those in \KPS\ for this parameter slice.  There is an analogous equality, up to the same factor of 2, between the function $a_{trial}$  for the magnetic duals.  It thus follows that the duality of \ILS\ for this product group has a non-empty superconformal window, which reduces to the $x$ interval of \KPS\ in this 1d subspace of the full parameter space of flavors and colors.  Likewise, all the dualities of \ILS\ have a non-empty superconformal window, for any $k$, which reduces to the $x$ interval of \KPS\ in a 1d subspace of the full parameter space of flavors and colors.

\listrefs

\end